\def\Ek{\overline{E}_k}
\def\Em{\overline{E}_m}
\def\eqn#1{\eqno(#1)}
\def\mi{\medskip\noindent}
\documentstyle[12pt,psfig]{article}
\begin{document}

\begin{center}
{\large\bf A ROUTE TO MAGNETIC FIELD REVERSALS:\\
AN EXAMPLE OF AN ABC-FORCED NON-LINEAR DYNAMO}

\bigskip
O.M. Podvigina\footnote{E-mail: olgap@mitp.ru}

\bigskip
International Institute of Earthquake Prediction Theory\\
and Mathematical Geophysics,\\
79 bldg.~2, Warshavskoe ave., 113556 Moscow, Russian Federation

\bigskip
Laboratory of General Aerodynamics, Institute of Mechanics,\\
Lomonosov Moscow State University,\\
1, Michurinsky ave., 119899 Moscow, Russian Federation

\bigskip
Observatoire de la C\^ote d'Azur,\\
BP~4229, 06304 Nice Cedex 4, France

\bigskip
Submitted to {\it Geophysical \& Astrophysical Fluid Dynamics}\\
26 January 2003
\end{center}

\begin{abstract}\noindent
We are investigating numerically the non-linear behaviour of a space-periodic
MHD system with ABC forcing. Most computations are performed for
magnetic Reynolds numbers increasing from 0 to 60 and a fixed
kinematic Reynolds number, small enough for the trivial solution with a
zero magnetic field to be stable to velocity perturbations. At the critical
magnetic Reynolds number for the onset of instability of the trivial solution
the dominant eigenvalue of the kinematic dynamo problem is real. In agreement
with the bifurcation theory new steady states with non-vanishing magnetic
field appear in this bifurcation. Subsequent bifurcations are investigated.
A regime is detected, where chaotic variations of the magnetic field
orientation (analogous to magnetic field reversals) are observed in the
temporal evolution of the system.

\end{abstract}

\pagebreak
\section{Introduction}

In numerical modelling of physical phenomena two approaches are feasible.
On the one hand, one can develop a model as detailed, as possible.
Then an obvious difficulty is encountered: usually simplifications are
necessary, owing to complexity of natural phenomena and/or lack of
precise data. Nevertheless, if the model is accurate enough, it
displays behaviour similar to that of the underlying physical system,
when fixed parameter values equal or close to those in the original system
are employed. On the other hand, one can consider a class of approximate
models, reflecting only global features of the underlying physical system.
Parameter values are varied in order to study how this affects
the overall behaviour of the system. The second approach is advantageous
in that properties characterising the class of systems are
revealed; computationally it is usually less demanding, than the first one.

We follow the second approach to examine a succession of
temporal regimes which can occur in magnetohydrodynamic (MHD) systems,
when Reynolds numbers are increased. In particular, we are concerned with
two questions: Which features of MHD systems are responsible for magnetic
field reversals -- an attribute of the Earth's magnetic field?
Which sequences of bifurcations can bring an MHD system to such a regime?

A comprehensive review of low-dimensional ODE dynamo models of magnetic field
reversals of the Earth, discussed in literature prior to 1994, can be found in
Jacobs (1994), Chapter 5. The models were categorised into two distinct classes.
In the models of the first class it was assumed that reversals were related
to MHD instabilities, triggered by finite-amplitude
perturbations of otherwise stable nonlinear dynamos. The perturbations were
supposed to be controlled by random processes, and various physical mechanisms
of non-MHD nature were proposed as their possible sources.
In models of the second class the reversals were a feature of the system,
with no external perturbations required to initiate a reversal.
The model of solar magnetic field reversals by Zeldovich {\it et al.}
(1983) also falls within this class. Hide {\it et al.} (1996) and Moroz
{\it et al.} (1998) generalised the Rikitake's disk model.

Armbruster {\it et al.} (2001) proposed a 7-dimensional ODE dynamo model
obtained by center manifold reduction with the use of symmetry considerations.
They considered convection in a spherical shell without rotation for such parameter
values, that mode interaction took place and heteroclinic cycles emerged
in the phase space, connecting the 7 equilibria present in their system.
The heteroclinic cycles were stable due to the symmetries of the system.
Introduction of magnetic field could transform the hydrodynamic equilibria
into MHD ones, with a non-vanishing magnetic field.
The resultant heteroclinic cycles involved changes of orientation of
the magnetic field, i.e., reversals of the magnetic dipole (non-dipole magnetic
components were ignored in their truncated system).

In the model of Melbourne {\it et al.} (2001) reversals were also
linked with an underlying heteroclinic cycle. To obtain the cycle, they
considered convection in a spherical shell with rotation and derived
a normal form near a codimension-3 bifurcation, assuming that
the flow sustaining the dynamo possessed a certain group of
symmetries. The normal form, truncated up to terms of the third order,
involved 9 parameters. When several higher-order small symmetry-breaking terms
were added in the normal form, for some parameter values the behaviour
of the system resembled the Earth's magnetic field reversals.

Reversals were observed in simulations of MHD systems with the geometry and
boundary conditions corresponding to those of the Earth in the
\hbox{3-dimensional} model (Glatzmaier and Roberts, 1995, 1996; Roberts and
Glatzmaier, 2001), involving the Navier-Stokes equation with the Coriolis,
buoyancy and Lorentz forces, the magnetic induction and
the temperature equations. A $2{1\over 2}$-dimensional model with a limited
resolution in the azimuthal direction, derived from these PDE's, also featured
magnetic reversals (Sarson and Jones, 1999; Sarson, 2000).
Magnetic field generated in these models was predominantly dipolar;
the lengths of time intervals of constant polarity and a short duration
of each reversal were consistent with those of the Earth.
However, the computations were very demanding in CPU time, prohibiting
numerical identification of complex bifurcations in such systems.

In simulations of Glatzmaier and Roberts reversals were robust.
Glatzmaier and Roberts (1996) used a modified set of equations
(compared to the one studied by Glatzmaier and Roberts, 1995:
compositional buoyancy was added and the anelastic approximation was used
instead of the Boussinesq approximation), but this did not result in significant
changes of the temporal behaviour of magnetic field. Despite the model was run
very far from the regime of the geophysical parameter values (Jones, 2000),
there was a strong similarity between the model output and the geodynamo.
This is another indication that their MHD attractor is structurally
stable (i.e., persists when equations, geometry, or parameter values are
considerably varied). Roberts and Glatzmaier (2001) considered three different
radii of the inner core, corresponding to its present size and some values
in the past and in the future (geodynamo was driven by the energy influx due
to the latent heat release at the freezing surface of the inner core, slowly
growing in time). Again, in all the three cases magnetic field of comparable
strength was generated, and the magnetic field exhibited a quantitatively similar
temporal behaviour, the system in the future displaying a stronger temporal
variability.

\section{Statement of the problem}

Equations of magnetohydrodynamics are invariant under the symmetry, which
preserves flows $\bf v$ and changes orientation of magnetic fields $\bf b$:
$$h:({\bf v},{\bf b})\to({\bf v},-{\bf b}).\eqn{1}$$
One can anticipate the following scenario of development of a temporal regime
involving magnetic field reversals. Suppose for certain values of a control
parameter the system possesses a stable steady state, in which magnetic field
vanishes. When the parameter is varied beyond the critical value,
two steady states with a non-vanishing
magnetic field emerge in a pitchfork bifurcation (generic to systems with the
symmetry (1)~). In subsequent bifurcations the steady states bifurcate into
more complex attractors, related by (1) and remaining so far distinct.
As the parameter is further varied, the attractors of the system join
(e.g., due to collision of attractors, or via heteroclinic connection)
and a unique attractor emerges. A sample trajectory on the attractor jumps
between former attractors (which degenerate into unstable invariant sets).
Since the jumps involve changes in the magnetic field orientation,
the resulting intermittency can thus be linked with the reversals.

The goal of the present study is to check whether this scenario is feasible.
As a test case we consider the Navier-Stokes equation
$${\partial{\bf v}\over\partial t}={\bf v}\times(\nabla\times{\bf v})
+{1\over R}\Delta{\bf v}-{\bf b}\times(\nabla\times{\bf b})
+{\bf f}-\nabla p\eqn{2.a}$$
and the magnetic induction equation
$${\partial{\bf b}\over\partial t}=\nabla\times({\bf v}\times{\bf b})
+{1\over R_m}\Delta{\bf b}\eqn{2.b}$$
under the solenoidality conditions
$$\nabla\cdot{\bf v}=0,\quad\nabla\cdot{\bf b}=0.\eqn{2.c}$$
Here $\bf v$ is velocity of the flow, $\bf b$ -- magnetic field, $p$ -- pressure,
$R$ and $R_m$ are kinematic and magnetic Reynolds numbers.
(Since in the problem under consideration characteristic length and velocity
magnitudes are of order one, the Reynolds numbers are defined
as inverse viscosity and inverse magnetic diffusivity, respectively.)
The fields are supposed to be $2\pi$-periodic in space.

The force
$${\bf f}={\bf u}_0/R\eqn{3}$$
is assumed, where ${\bf u}_0$ is an ABC flow
$${\bf u}_0=(A\sin x_3+C\cos x_2,\ B\sin x_1+A\cos x_3,
\ C\sin x_2+B\cos x_1).\eqn{4}$$
For this force,
$${\bf v}={\bf u}_0,\quad{\bf b}=0\eqn{5}$$
is a steady solution to (2)-(4) for all Reynolds numbers.

This particular system is considered for the following reasons:
First, space periodicity enables one to use pseudo-spectral
methods (see Canuto {\it et al.}, 1989; Boyd, 1989),
which are computationally efficient.
Second, linear stability of the system (2)-(5) to
hydrodynamic (Galloway and Frisch, 1987;
Podvigina and Pouquet, 1994; Podvigina, 1999) and magnetic
(Arnold and Korkina, 1983; Galloway and Frisch, 1986; Galanti {\it et al.}, 1992;
Childress and Gilbert, 1995) perturbations has been investigated
both numerically and analytically (nonlinear regimes for some parameter
values were also explored by
Podvigina and Pouquet, 1994; Podvigina, 1999; Galanti {\it et al.}, 1992;
Feudel {\it et al.}, 1996; Brummell {\it et al.}, 2001).
These studies provide guidance, in what parameter ranges
the targeted type of behaviour may be observed.
In particular, growing magnetic modes exist for the flow (3)
in broad intervals of constants $A$, $B$ and $C$. Third,
since the critical magnetic Reynolds number is typically of the order of 10,
only moderate $R_m$ need to be considered, and
a relatively low resolution ($32^3$ Fourier harmonics) suffices.

Magnetic field growth rates were calculated by Galanti {\it et al.} (1992)
for the flow (4) with the coefficients satisfying
$B=C,\ A^2+B^2+C^2=3,\ 0<B/A<1$
for various values of $R_m$. For $R_m=12$ there are three windows in $B/A$
of positive growth rates (see Fig.~2 ibid.).
We have checked that in the window $0.7<B/A<0.9$ the dominant eigenvalue is real,
and in the other two windows it is complex.
Distinct steady states with a non-zero magnetic field
appear in a bifurcation of the trivial steady state (5), only if
the magnetic induction operator has a zero (hence real) eigenvalue
at the bifurcation point. This has suggested to set
$$A=1,\quad B=C=0.75;\eqn{6}$$
these values of constants in (4) are assumed throughout.

With our goal in mind, in the present study we focus at non-hydrodynamic attractors.
The hydrodynamic global stability of ${\bf u}_0$ guarantees that magnetic
field does not vanish in saturated regime, if ${\bf v}={\bf u}_0$ is a
kinematic dynamo. We have verified that for $R=4$, ${\bf v}={\bf u}_0$ is
a unique attractor of the hydrodynamic system (~(2.a,c) with ${\bf b}=0$),
and most computations are made for this value of $R$.
Several runs are also performed for higher $R$.

\section{The group of symmetries}

The group of symmetries without inversion of time of an ABC flow
with $B=C$, which we denote by $\cal G$, is comprised of two independent symmetries
$$s_1:\ x_1\to-{\pi\over2}-x_3,\ x_2\to-{\pi\over2}-x_2,\ x_3\to-{\pi\over2}-
x_1,$$
$$s_2:\ x_1\to{\pi\over2}+x_3,\ x_2\to-{\pi\over2}+x_2,\ x_3\to{\pi\over2}-
x_1,$$
their superpositions
$$s_3=s_2^2:\ x_1\to\pi-x_1,\ x_2\to\pi+x_2,\ x_3\to-x_3,$$
$$s_4=s_2^3:\ x_1\to{\pi\over2}-x_3,\ x_2\to{\pi\over2}+x_2,\ x_3\to-
{\pi\over2}+x_1,$$
$$s_5=s_1s_2:\ x_1\to\pi+x_1,\ x_2\to-x_2,\ x_3\to\pi-x_3,$$
$$s_6=s_1s_2^2:\ x_1\to-{\pi\over2}+x_3,\ x_2\to{\pi\over2}-x_2,\
x_3\to{\pi\over2}+x_1$$
$$s_7=s_1s_2^3:\ x_1\to-x_1,\ x_2\to\pi-x_2,\ x_3\to\pi+x_3,$$
and the identity transformation $s_8=e$ (Arnold, 1984; Dombre {\it et al.},
1986; Podvigina and Pouquet, 1994). The group is isomorphic
to the symmetry group of a square, ${\bf D}_4$.

Any symmetry of the force (3), (4) which does not involve inversion of time
is also a symmetry of the Navier-Stokes equation (2.a) with ${\bf b}=0$.
The group of symmetries of the MHD system (2)-(4), (6),
which we denote by $\cal H$, is a direct product of
$\cal G$ and the 2-element group ${\bf Z}_2$, generated by $h$ (1).
It has 16 elements, which are either $s_i$, or $hs_i$.

Problems in hydrodynamics and magnetohydrodynamics (see e.g. the review
by Crawford and Knobloch, 1991) often involve symmetries, induced by
the geometry of the region where fluid is contained. In particular,
if convection in a plane layer is restricted to space-periodic flows
with equal periods along two Cartesian axes parallel to boundaries,
and conditions at the boundaries are different,
the group of symmetries of the system is isomorphic to $\cal G$.
If in addition magnetic field is present, the symmetry group
is isomorphic to $\cal H$. Thus, though the system under consideration
(2)-(4), (6) is highly symmetric, it can not be regarded as an exotic one.

Symmetries are useful for categorisation of attractors.
Let $\cal A$ be an attractor of a dynamical system, invariant under
a symmetry $g$: $g({\cal A})=\cal A$. Two cases can be distinguished: either
$\cal A$ is pointwise invariant, i.e.~$g({\bf x})={\bf x}$ for all
points ${\bf x}\in\cal A$, or it is invariant only as a set, with
$g({\bf x})\ne{\bf x}$ for some \hbox{${\bf x}\in\cal A$}. In what follows,
only the symmetries for which an attractor is pointwise invariant
are regarded as symmetries of the attractor.

\section{Attractors of the MHD system for $R=4$}

\begin{table}
Table 1.
Attractors, detected for the MHD system with the forcing (3), (4), (6)
for $R=4$ and $0\le R_m\le 60$. The third column shows the number of elements of
the symmetry group for which an attractor is pointwise invariant,
and the fourth column -- generators of the group.
Indices in the labels of attractors (second column) have the following meaning.
For periodic orbits emerging in the period doubling cascade the superscript
$M$ refers to an $M\tau$-period orbit, $\tau$ being the period of the foremost
orbit. The first subscript labels different attractors of a given kind.
The second subscript labels mutually symmetric attractors.

\bigskip\hspace*{-2cm}
\begin{tabular}{|c|c|c|c|c|c|}\hline
&&Number&&&\\
$R$&Attractors&of sym-&Generators&$\Ek$&$\Em$\\
&&metries&&&\\\hline
$R_m\le 14$&steady state $S_0$&16&$s_1$, $s_2$, $h$&1.06&0\\\hline
$15\le R_m\le 24$&steady states $S_{1,i}$ ($i=1,2)$&8&$hs_1$, $s_2$&
1.06-0.92&0-0.035\\\hline
$25\le R_m\le 37$&periodic orbits $P_{1,i}$ ($i=1,2)$&4&$hs_1$,
$s_3(=s_2^2)$&0.94&0.03\\\hline
$37.5\le R_m\le 38.5$&steady states $S_{1,i}$ ($i=1,2$)&8&$hs_1$, $s_2$&0.95&
0.025\\\hline
$39\le R_m\le 39.2$&periodic orbits $P_{2,i}^1$ ($i=1,2$)&8&$hs_1$,
$s_2$&0.98&0.02\\\hline
$39.3\le R_m\le 39.4$&periodic orbits $P_{2,i}^2$ ($i=1,2$)&8&$hs_1$,
$s_2$&0.98&0.02\\\hline
$R_m=39.45$&periodic orbits $P_{2,i}^4$ ($i=1,2)$&8&$hs_1$, $s_2$&0.98&0.02\\\hline
$R_m=39.5$&chaotic $C_{1,i}=P_{2,i}^{\infty}$ ($i=1,2)$&8&$hs_1$, $s_2$&0.98&
0.02\\\hline
$R_m=39.6$&periodic orbits $P_{2,i}^1$ ($i=1,2$)&8&$hs_1$, $s_2$&0.97&0.025\\
&\& periodic orbit $P_3$&8&$hs_1$, $s_2$&0.95&0.03\\\hline
$39.7\le R_m\le 41$&periodic orbit $P_4$&8&$hs_1$, $s_2$&0.93&0.035\\\hline
$R_m=42,43,44$&tori $T_{1,i}$ ($i=1,2$)&4&$hs_1$, $s_3$&0.93&0.04\\\hline
$R_m=45$&chaotic $C_{2,i}$ ($i=1,2$)&4&$hs_1$, $s_3$&0.93&0.04\\\hline
$R_m=46$, 47&chaotic $C_{3,i}$ ($i=1,4$)&2&$hs_1$&0.94&0.03\\\hline
$R_m=48$, 49&chaotic $C_{4,i}$ ($i=1,4$)&2&$hs_1$&0.92&0.035\\\hline
$R_m=50$, 51&chaotic $C_{5,i}$ ($i=1,2$)&2&$hs_1$&0.91&0.04\\\hline
$52\le R_m\le 58$&tori $T_{2,i}$ ($i=1,4$)&2&$hs_1$&0.9&0.05\\\hline
&\vspace*{-5.5mm}&&\\\hline
$16\le R_m\le 25$&periodic orbit $P_6$&4&$hs_2$&0.4-0.7&0.07-0.2\\\hline
$26\le R_m\le 60$&torus $T_3$&4&$hs_2$&0.3-0.4&0.2-0.25\\\hline
\end{tabular}
\end{table}

\bigskip
Results of computations are summarised in Table 1 (see also the bifurcation
diagram Fig.1). The magnetic Reynolds number has been increased step 1,
except in the interval $37<R_m<40$, where a high density of bifurcations
has required smaller steps.
For every considered $R_m$ an attractor has been obtained in a run
with an initial condition, which is a perturbation of the trivial steady state
(5) with the energy of $10^{-6}$ in each Fourier harmonics spherical shell.
The range of existence of attractors has been determined by continuation in
parameter: the runs are done with an initial condition, which
is a point on the attractor for close values of $R_m$.

For $15\le R_m\le 58$ the system under consideration possesses multiple
attractors. For \hbox{$16\le R_m\le 58$} attractors, which we have detected, belong
to two independent families; within each family they are genetically related
by sequences of bifurcations. Attractors from different families
have distinct groups of symmetries. The families can be distinguished
also by their time-averaged kinetic ($\Ek$) and magnetic ($\Em$) energies:
$0.9<\Ek<1.1$, $0\le\Em<0.05$ for the first family,
and $0.3<\Ek<0.7$, $0.07<\Em<0.25$ for the second one.

The first family exists for $0\le R_m\le58$: the first attractor
is the trivial steady state (5) (denoted by $S_0$) remaining
stable up to\footnote{This is consistent with results of
Galanti {\it et al.} (1992): since the
ABC coefficients (6) are not normalised to satisfy $A^2+B^2+C^2=3$ as they were
in the cited paper, Reynolds numbers referred to here are lower
by a factor $(17/24)^{1/2}$.} $R_m=14$.
It becomes unstable in a pitchfork bifurcation, in which two stable mutually
symmetric steady states ($S_{1,1}$ and $S_{1,2}$) with a non-zero magnetic
field emerge. The \hbox{8-element} symmetry group of $S_{1,i}$
is isomorphic to ${\bf D}_4$; however, it is distinct from $\cal G$.

The steady states $S_{1,i}$ become unstable in a supercritical Hopf bifurcation
at the interval $24<R_m<25$. A stable periodic orbit $P_{1,i}$ of period
$\approx 12$ appears in a vicinity of the steady state
$S_{1,i}$. The two orbits are mutually symmetric (they are interrelated by
$h$~). Each periodic orbit possesses all 8 symmetries of its parent steady
state, but each individual point of the orbit
has a symmetry group of only 4 elements (it is isomorphic to ${\bf D}_2$).
The orbits exist for $25\le R_m\le 37$ remaining attracting.
A typical behaviour of a trajectory is shown on Fig.~2 for
$R_m=30$. After an initially small magnetic energy at first exponentially
grows and afterwards decays with oscillations, it subsequently levels off
(see a plateau about 400 time units long on Fig.~2a): the trajectory
in the phase space evolves in a vicinity of $S_{1,i}$ (which is unstable now).
In the further evolution the trajectory leaves the steady state, being
attracted by $P_{1,i}$ (see Fig.~2b). The orbits disappear in a subcritical
Hopf bifurcation between $R_m=37$ and $R_m=37.5$~. For $R_m=37.5$, 38 and 38.5
the two steady states $S_{1,i}$ are verified to be stable.

Next bifurcation of $S_{1,i}$ is again a supercritical Hopf bifurcation.
The emerging mutually symmetric periodic orbits $P_{2,i}^1$, $i=1,2$
(one for each of the two steady states) possess all the symmetries
of the steady states. For $39\le R_m\le 39.2$ they are attracting,
and their period is $\tau(P_{2,i}^1)\approx 120$. Behavior of a sample
trajectory in the phase space (see Fig.~3) resembles the one shown on Fig.~2;
however, on Fig.~3a the plateau corresponding to the evolution in the vicinity
of the unstable steady state $S_{1,i}$ virtually disappears, and the period
is much larger. The orbit is located in a different region of the phase space
(cf.~Fig.~2b and Fig.~3b). At an $R_m$ between 39.2 and 39.3 two mutually
symmetric orbits $P_{2,i}^2$ ($i=1,2$) of a twice larger period emerge
(see Fig.~4a). A sequence of period-doubling bifurcations begins.
The next period doubling occurs at the interval $39.4<R_m<39.45$
(see period-four orbit $P_{2,i}^4$ for $R_m=39.45$ on Fig.~4b).
At $R_m=39.5$ a trajectory, initially close to $S_0$,
is either already chaotic, or close to an orbit of a very long period
$P_{2,i}^\infty$ (see Fig.~4c). This indicates that at $R_m=39.5$ the
period-doubling cascade is over. For $R_m=39.6$ a trajectory
with the same initial condition is attracted to a period-one orbit $P_{2,i}^1$.

Another attractor exists for $R_m=39.6$~. It reveals itself when the
attractor for $R_m=39.7$, which is unique in this region of the phase space
(see below) is continued in smaller $R_m$ (no attractor of this type is observed
by further continuation in smaller $R_m$). This is a periodic orbit $P_3$,
involving transitions between the two period-two orbits $P_{2,1}^2$ and
$P_{2,2}^2$ (see Fig.~5a). The orbit is invariant under $h$ and any symmetry
from $\cal G$. In the course of temporal evolution,
direction of magnetic field is reversed, behaviour of non-zero
magnetic field Fourier coefficients is similar to that shown on Fig.~5b.
A reversal takes a relatively short time, during which initially
the energy of magnetic field attains the minimum and kinetic -- the maximum,
and subsequently magnetic energy blows up to a maximum,
with the kinetic energy simultaneously reaching its minimum (Fig.~5c,d).
Thereby, during a reversal magnetic energy is at first transformed into kinetic
one, and then back into magnetic energy; for higher Reynolds numbers reversals
exhibit such energy transformations as well.

This bifurcation is apparently of the following nature.
In the period-doubling cascade infinitely many unstable periodic orbits
$P_{2,i}^M$ ($i=1,2$) of periods $\tau(P_{2,i}^M)=M\tau(P_{2,i}^1)$ are created.
Consider two such orbits $P_{2,i}^m$ ($m=2$ in our case) related by $h$.
Suppose that for some value of a control parameter the stable manifold of
$P_{2,1}^m$ intersects the unstable manifold of $P_{2,2}^m$,
and a heteroclinic trajectory connecting these orbits emerges.
Consider the symmetry, which maps one of the periodic orbits to another.
Under this symmetry the heteroclinic trajectory is mapped to a distinct
heteroclinic trajectory, and the two form a heteroclinic cycle.
The cycle may give rise to a periodic orbit $P_3$, when
the control parameter varies. (This scenario is analogous to that in
two-dimensional dynamical systems, where a homoclinic connection generically
implies for close values of the control parameter existence of periodic
orbits in a vicinity of the homoclinic trajectory, and their period tends to
infinity when the critical value is approached.) Thus, the regime of reversals
emerges via heteroclinic connection of two periodic orbits.

For $39.7\le R_m\le 41$ sample trajectories, initially close
to the trivial steady state $S_0$, are attracted by a periodic orbit $P_4$.
While $P_3$ (for $R_m=39.6$)
follows closely each of $P_{2,i}^2$ (Fig.~5a), this is not the case any more
for $P_4$ (for the higher $R_m$'s; see Fig.~6). In agreement with our conjecture
that emergence of the orbit is related to heteroclinic connection,
the period decreases from 800 for $R_m=39.7$ to 500 for $R_m=41$.

For $39.6\le R_m\le 41$ the attractor involving reversals is unique.
Behaviour of all magnetic field Fourier harmonics is coherent and resembles
the one shown on Fig. 5b: time intervals, where a Fourier coefficient remains
close to a certain value, are separated by jumps changing the sign of
the coefficient, but not its magnitude. In contrast with the coherent
behaviour of all magnetic field Fourier harmonics for $39.6\le R_m\le 41$,
they can be divided for \hbox{$42\le R_m\le 51$} into two groups
with a qualitatively different behaviour.
Behaviour of harmonics from the first group is the same as above, and hence
the system exhibits reversals. Harmonics from the second group oscillate
with a relatively short temporal period, some of their time averages
do not vanish. The total time-averaged energy of harmonics from the first
group is higher than that from the second group (for instance, cf. Fig.7a
and 7b). The Earth's magnetic field shows a similar behavior: Spherical
harmonics can be categorized into the dipole and the quadrupole families
according to their symmetry about the equatorial plane (Roberts and Stix, 1972),
and paleomagnetic data suggest that a magnetic reversal in the Earth is more
likely to occur, when the ratio of the magnitude of the quadrupole-family
component of the magnetic field to that of the dipole-family component is high
(Merrill and McFadden, 1988).
For $42\le R_m\le 51$ our attractors have no the symmetry $h$ on average
(since the time average of some harmonics is non-zero), and thus $h$ maps
an attractor to a different attractor with reversals. Again, a parallel can be drawn
with a property of the magnetic field of the Earth: Paleomagnetic data
indicate that the time-averaged magnetic field of the Earth lacks the north-south
symmetry and there are significant (though small) differences between the states
of the normal and reverse polarity (Merrill {\it et al.}, 1979), this implying
the absence of the symmetry $h$ on average.

For $R_m=42$ and 43 the regime is quasi-periodic: the orbit $P_4$ bifurcates
into two attracting tori $T_{1,i}$ ($i=1,2$) interrelated by $h$ (see Fig.~7).
The second frequency can be observed on Fig.~7a, showing real part of the
Fourier component $b^3_{0,1,1}$ of magnetic field. The time average of
this Fourier component is negative (Fig.~7a); hence, the symmetry $h$ maps
this attractor to a distinct one. (For $37.5\le R_m\le41$ the component vanishes,
because former attractors [$P_4$ for $39.7\le R_m\le41$, $P_3$ for $R_m=39.6$,
$P_{2,i}^m$ for $39\le R_m\le39.6$ and $S_{1,i}$ for $37.5\le R_m\le38.5$]
possess the symmetry $hs_5=hs_1s_2$.)

For $R_m=45$ two mutually symmetric attractors persist (Fig.~8):
each torus $T_{1,i}$ bifurcates into a chaotic attractor $C_{2,i}$.
Reversals are less regular, yet they are
too ordered in comparison with those of the Earth's magnetic field.

For $R_m=46$ (Fig.~9) behaviour of a sample trajectory suggests existence of
four (unstable) steady states $S_{2,i}$ (see plateaux
at $1500\le t\le2100$). They are mutually symmetric, and have the same
4-element symmetry group generated by $hs_1$ and $s_3$. Apparently
two $S_{2,i}$'s emerged from each $S_{1,i}$ in a pitchfork bifurcation.
Initially (at $0\le t<700$) the sample trajectory for $R_m=46$
undergoes several reversals, similar in nature to those
observed for \hbox{$39.7\le R_m\le45$}. Afterwards it is attracted to a
steady state $S_{2,i}$.
The behaviour near the steady state resembles the Shilnikov attractor:
the trajectory approaches the steady state along the two-dimensional
eigenspace associated with the complex eigenvalues with the
maximal negative real part, and leaves it along the one-dimensional
unstable manifold. Subsequent sample evolution consists of
transitions between the steady state $S_{2,i}$ and the region of the phase space
where a former chaotic attractor $C_{2,i}$ was located (duration
of repeating events in the saturated regime is $\approx 1700$).
In particular, large-amplitude excursions at
$2350\le t\le2500$, $2300\le t\le2450$, $4000\le t\le4150$, etc., are
reminiscent of the behaviour of a $C_{2,i}$ trajectory for $R_m=45$.
Thus, apparently two former chaotic attractors $C_{2,i}$
have disappeared in collision with the four steady states
$S_{2,i}$ to give rise to four new chaotic attractors $C_{3,i}$.
It is notable that the sample trajectory leaves a vicinity of the $S_{2,i}$
in alternating directions along the unstable manifold.
For $R_m=47$ the behaviour is similar to the one at $R_m=46$.

For $R_m=48$ the system possesses four mutually symmetric attractors $C_{4,i}$.
Events are longer than in the previous regime, about 3000 time units
(see Fig.~10). They exhibit a new feature, a phase of initially
exponentially decaying oscillations (e.g. at $1700\le t\le2300$ and
$4500\le t\le 5200$). This indicates existence of
a weakly unstable periodic orbit of a new kind, $P_{5,i}$ ($i=1,4$),
in a vicinity of the former chaotic attractor $C_{3,i}$.
A similar behaviour is also observed for $R_m=49$.

For $R_m=50,51$ (Fig.~11) the four $C_{4,i}$ are superceded
by two new attractors, $C_{5,i}$. Large amplitude oscillations, e.g. at
$7500\le t\le 8500$, are due to attraction of the trajectory by weakly
unstable tori $T_{2,i}$ ($i=1,4$), which have bifurcated from the periodic
orbits $P_{5,i}$. In the saturated regime behaviour consists of three phases:
($i$) a trajectory is close to a periodic orbit $P_{5,i}$
(e.g. at $2300\le t\le 3000$ and $4300\le t\le 5000$); ($ii$) the trajectory
evolves in the vicinity of a torus $T_{2,i}$ (e.g. at $3200\le t\le 4000$ and
$5300\le t\le 6000$); ($iii$) the trajectory abruptly jumps toward the second
periodic orbit $P_{5,i'}$ to reproduce the sequence of phases.

For $R_m=52$ each of the two former chaotic attractors splits into two
attracting tori $T_{2,i}$ (Fig.~12; cf.~Figs.~11b and 12b).
The four new attractors are mutually symmetric,
they are stable for $52\le R_m\le 58$.

For $57\le R_m\le60$ a sample trajectory initially close to $S_0$ is attracted
by the torus $T_3$, belonging to the second family.
The family emerges between $R_m=15$ and $R_m=16$: in addition to the
steady states $S_{1,i}$, for $R_m\ge 16$ the system possesses another attractor,
a stable periodic orbit $P_6$. The orbit does not exist for $R_m\le15$;
apparently it appears in a saddlenode bifurcation. For $R_m=26$ the orbit is
unstable and a torus $T_3$ has appeared in a Hopf bifurcation. $T_3$ remains
an attractor for all higher considered $R_m$ (specifically, for $R_m=26$,
27, 30, 40, 57 and 60). For $R_m\ge 57$ the torus attracts trajectories
(in the phase space) initially close to the trivial steady state $S_0$, and for
\hbox{$R_m\ge 59$} it is the only global attractor of the system.
The torus $T_3$ and the orbit $P_6$ were traced back from $R_m=57$ by
continuation in $R_m$.

Studying numerically a dynamical system, one can never guarantee that all
attractors are found. The computations are intensive (a run with
the resolution $32^3$ Fourier harmonics up to $t=5000$ has
required several dozens hours of one processor), prohibiting
to carry out more detailed computations. For instance, if attractors of two
or more new types emerge for a particular considered $R_m$, in principle
we could have missed some of them; however, it is unlikely that several attractors
of different morphology are born at the same or close values of the magnetic
Reynolds number. In our estimations all attractors, whose basin of attraction
contains the trivial steady state, have been identified -- up to a natural
limitation: since a period-doubling cascade involves an infinite number of
bifurcations, it is impossible to capture numerically {\it all} bifurcations
within the cascade. Attractors, which do not attract small perturbations
of the trivial steady state and which are far from it in the phase space,
(if such attractors exist in our system) are not involved in the scenario
of emergence of magnetic field reversals, and they are out of the scope
of the present investigation.

\bigskip
\section{Attractors of the MHD system for $R_m=40$}

\bigskip
Simulations are performed for $R_m=40$ and for several values of $R$
with an initial condition, same for all runs, being a small perturbation of
$S_0$ (see Table~2). The question we address is how the temporal behaviour
changes with $R$, in particular, if for higher $R$ reversals persist, or whether
magnetic field vanishes in a saturated regime. We do not attempt to identify
all bifurcarions occurring when $R$ varies.
Computations with the same $R$ and the same initial condition
for the flow are also performed for the purely hydrodynamic system (2.a),
in order to compare the behaviour and attractors of the two systems.
(If (2) possesses attractors with ${\bf b}=0$ different from
the trivial steady state, an initially growing magnetic field can
die out in the nonlinear regime, the resulting attractor being purely
hydrodynamical, see Brummell {\it et al.}, 1998; Matthews, 2000.)

\begin{table}
Table 2. Attractors, detected for the MHD system with the force (3), (4), (6)
for $R_m=40$ and $3\le R\le 25$. The third column shows the number of elements
of the symmetry group for which an attractor is pointwise invariant,
and the fourth -- generators of the group.

\bigskip\noindent
\begin{tabular}{|c|c|c|c|c|c|}\hline
$R$&Attractors&Number of&Generators&$\Ek$&$\Em$\\
&&symmetries&&&\\\hline
$R=3$ & $T_3$ & 4 & $hs_2$ & 0.4 &0.25\\\hline
$R=4$&periodic orbit $P_4$ &8&$hs_1$, $s_2$ & 0.95 &0.03\\\hline
$R=6$&chaotic $C_{2,i}$ ($i=1,2$)&4&$hs_1$, $s_3$ & 0.95 &0.03\\\hline
$R=10$ &periodic orbit $P_7$ &8&$hs_1$, $s_2$ &0.5 &0.06\\\hline
$R=15$ &chaotic $C_6$ &4&$s_1$, $s_3$ & 0.6 &0.04\\\hline
$R=20$ &tori $T_{4,i}$ ($i=1,2$) &4&$s_1$, $s_3$ & 0.7 &0.01\\\hline
$R=25$ &chaotic $C_{7,i}$ ($i=1,2$)&1&$e$ & 0.5 &0.02\\\hline
\end{tabular}

\vskip1cm
Table 3.
Attractors, detected for the hydrodynamic system (2.a) with the force (3),
(4), (6) for $3\le R\le 25$. The third column shows the number of elements of
the symmetry group for which an attractor is pointwise invariant,
and the fourth column -- generators of the group.

\bigskip\noindent
\begin{tabular}{|c|c|c|c|c|}\hline
$R$&Attractors&Number of&Generators&$\Ek$\\
&&symmetries&&\\\hline
$R=3,4,6$ & ${\bf u}_{ABC}$ & 8 & $s_1$, $s_2$ &1.06\\\hline
$R=10^*$ & ${\bf u}_{ABC}$ & 8 & $s_1$, $s_2$ &1.06\\
& $S^h_1$ & 8 & $s_1$, $s_2$ &0.99\\\hline
$R=15$ & $S^h_1$ & 8 & $s_1$, $s_2$ &0.93\\\hline
$R=20$&chaotic $C_1^h$ &4&$s_2$ &0.9\\\hline
$R=25$&chaotic $C_2^h$ &1&$e$ &0.7\\\hline
\end{tabular}

\vskip6mm
\noindent
$^*$ $S^h_1$ for $R=10$ was traced back from $R=15$; it is not observed for
$R=6$.
\end{table}

For $R=3$ the detected attractor is the torus $T_1$ (which is an attractor
for $R=4$ and $26\le R_m\le60$). For $R=6$ reversals take place, similar
to those observed for $R=4$ and $R_m=45$, and the attractor is $C_{2,i}$.
Thus, for $R$ close to 4 no new attractors are found.

For $R=10$ a new attractor was found in the full MHD system --
a periodic orbit $P_7$ with a symmetry group of 8 elements.
Comparison of Tables 2 and 3 reveals no relation of attractors of
the hydrodynamic and MHD systems for $R=10$, 15 and 20.
For these Reynolds numbers a small magnetic field
(magnetic energy $E_m$ is below 0.07; see Table~2)
drastically changes behaviour of the system --
the hydrodynamic and MHD systems have attractors of different types, and
the average kinetic energy $\Ek$ decreases significantly, e.g.
$\Em=0.01$ and $\Ek$ decreases from 0.9
(the hydrodynamic case) to 0.7 (the MHD case) for $R=20$.

For $R=15$ behaviour of the sample trajectory of the MHD system
in saturated regime is chaotic. But before the saturated regime sets
in, the trajectory is attracted by $P_7$ and for $1000\le t\le 2000$
remains close to this periodic orbit (see Fig.~13), which is now weakly
unstable. For $R=20$ the trajectory with the same initial condition
is attracted to a new torus $T_{4,i}$; initially the temporal behaviour
of the trajectory resembles the one observed for $R=15$ (cf.~Figs.~14 and 14).

For $R=25$ the MHD system possesses new chaotic attractors $C_{7,i}$ \hbox{($i=1,2$)}
with a trivial symmetry group. They resemble the chaotic attractor $C_6$
observed for $R=15$ (cf.~Fig.~13 and Fig.~15), but unlike $C_{7,i}$,
$C_6$ is unique and has a symmetry group of four elements.
Comparison of $C_{7,i}$ with the attractor of the non-magnetic
Navier-Stokes equation, which also is chaotic, reveals that
the influence of magnetic field is in some sense stabilising:
Fourier coefficients of the flow experience fewer jumps
and the amplitude of their oscillations is much smaller
(compare Fourier components of flows on Figs.~15b and 16b).
An exponential growth of the initially small magnetic field in the sample
evolution of the system begins only at $t=200$ (Fig.~15c). Accordingly,
until $t=300$ the flow evolution is similar to the one in the absence of
magnetic field. Magnetic field starts growing, when the trajectory
is attracted by a weakly unstable periodic orbit in the hydrodynamic
subspace. Departure from this orbit causes an initial decay of magnetic
field before the onset of saturated behaviour.

For $R=15$ and 20 behaviour of Fourier components of magnetic field is
similar to the one shown on Fig.~15d. It consists of chaotic
irregular small-period oscillations about zero, different from the
behaviour observed for $R=4$. Thus for $R_m=40$, $15\le R\le25$ the considered
system is not in a regime resembling reversals.

\section{Conclusion}

The bifurcation scenario leading to emergence of reversals, which we put
forward, proves feasible. Reversals are found for small $R$,
for which the hydrodynamic system has a unique globally stable steady state.
Complexity of the sequence of bifurcations obtained in simulations
is comparable to that of the hydrodynamic system examined by
Podvigina and Pouquet (1994) and Podvigina (1999).
Numerous types of behaviour are identified, which involve a large variety of
attractors and unstable invariant sets of various types affecting
behaviour of evolutionary solutions. The regime of reversals arises as a result
of merging of two distinct attractors, apparently via heteroclinic connection
due to intersection of stable and unstable manifolds of periodic orbits
(in contrast to the system of Armbruster {\it et al.} (2001), where
a heteroclinic cycle emerged simultaneously with magnetic steady states).

Other regimes with a non-zero magnetic field are observed when
any of the Reynolds numbers is increased. Thus in our computations reversals are
not so robust, as in simulations of Glatzmaier and Roberts.
The range of parameters considered here is rich in bifurcations:
for $R=4$ we found 14 bifurcations in the interval $14<R_m<60$ (counting
the infinite number of period-doubling bifurcations as one),
7 distinct types of attractors involve magnetic field reversals.
In another dissimilarity with the results of Glatzmaier and Roberts (and with
the Earth's magnetic field) in our simulations \hbox{$\Em<\Ek$}
(cf. $\Em/\Ek\sim10^4$ for the geodynamo).
In our computations reversals appear more regular than those in the Earth,
since much smaller values of Reynolds numbers are considered.
The dissimilarity with reversals in the Earth also owes
to the fact that we solve a simplified system of PDE's
(neither an account of thermal or sedimentation-driven convection is taken,
nor that of the Coriolis force) in an idealised space-periodic geometry.
Nevertheless, it appears possible to mimick the presence of two time scales
(average periods of constant polarity are much longer than the average time
of a reversal itself); instances of excursions of magnetic field (i.e.,
incomplete reversals) can be observed as well (see e.g. Figs. 8a and 10).

Our results show that magnetic field reversals are an inherent
feature of a nonlinear dynamical system of the MHD type. The fundamental reason
for their existence is the symmetry $h$, changing polarity of the magnetic field.
Neither external triggering random processes, such as the ones postulated
in earlier studies, nor symmetry breaking due to inhomogeneity of the outer core
boundaries, nor rotation, nor a specific heteroclinic structure or symmetries
of the flow sustaining a non-linear dynamo are required to give rise to
a sequence of reversals. Such physical processes, as heat transfer or
sedimentation are also unnecessary {\it per se} (as soon as the system
is under the action of any body force providing input of energy).

\bigskip
{\bf Acknowledgments}

\bigskip
I am grateful to V.~Zheligovsky for discussions.
Part of this research was carried out during my stays
at the School of Mathematical Sciences, University of Exeter, UK,
in January -- February 2001 and in May -- July 2002.
I am grateful to the Royal Society for their support of the two visits.
Some numerical results were obtained using
computational facilities provided by the program ``Simulations
Interactives et Visualisation en Astronomie et M\'ecanique (SIVAM)''
at Observatoire de la C\^ote d'Azur, France. My research work at
Observatoire de la C\^ote d'Azur was supported by the French
Ministry of Education. Part of the work was done during a visit
to Departamento de Matem\'atica Aplicada, Faculdade de Ci\^encias,
University of Porto, Portugal at the kind invitation
of Centro de Matem\'atica Aplicada in June -- July, 2001.
Last but not least, I am grateful to referees for stimulating remarks.

\pagebreak
{\bf References}

\mi Armbruster D., Chossat P., and Oprea I.
Structurally stable heteroclinic cycles and the dynamo dynamics,
In: {\it Dynamo and Dynamics, a Mathematical Challenge}
(Eds. Chossat, P., Armbruster, D. and Oprea, I.), 313--322,
Kluwer Academic Publishers (2001).

\mi Arnold V.I. and Korkina E.I. The growth of a magnetic field
in a three-dimensional incompressible flow, {\it Vestnik Moscow State Univ.,
Ser.~Math.} {\bf 3}, 43--46 (1983) (in Russian).

\mi Arnold V.I. On the evolution of magnetic field under the
action of advection and diffusion, in {\it Some problems of modern
analysis} (ed. V.M. Tikhomirov). Moscow Univ. Press, 8--21 (1984) (in Russian).

\mi Boyd J.P. {\it Chebyshev \& Fourier Spectral
Methods}. Springer-Verlag, Berlin (1989).

\mi Brummell N.H., Cattaneo F. and Tobias S.M. Linear and nonlinear dynamo
action, {\it Phys. Lett.} A {\bf 28}, 437--442 (1998).

\mi Brummell N.H., Cattaneo F. and Tobias S.M. Linear and nonlinear dynamo
properties of time-dependent ABC flows, {\it Fluid Dyn.~Res.} {\bf 28},
237--265 (2001).

\mi Canuto C., Hussaini M.Y., Quarteroni A. and Zang T.A. {\it Spectral
Methods in Fluid Dynamics}, Springer-Verlag, Berlin (1988).

\mi Childress S. and Gilbert A.D. {\it Stretch, twist, fold: the fast dynamo}.
Springer-Verlag, Berlin (1995).

\mi Crawford J.D. and Knobloch E. Symmetry and symmetry-breaking bifurcations
in fluid dynamics, {\it Annu. Rev. Fluid Mech.} A {\bf 23}, 341--387 (1991).

\mi Dombre T., Frisch U., Greene J.M., H\'{e}non M., Mehr A. and
Soward A. Chaotic streamlines in the ABC flows, {\it J.~Fluid Mech.}
{\bf 167}, 353--391 (1986).

\mi Feudel S., Seehafer N., Galanti B. and Schmidtmann O. Symmetry
breaking bifurcation for the magnetohydrodynamic equation with helical
forcing, {\it Phys.~Rev.~E} {\bf 54}, 2589--2596 (1996).

\mi Galanti B., Sulem P.L. and Pouquet A. Linear and Non-Linear
Dynamos Associated With ABC Flows, {\it Geophys.~Astrophys.~Fluid Dynamics}
{\bf 66}, 183--208 (1992).

\mi Galloway D.J. and Frisch U. Dynamo action in a family of
flows with chaotic streamlines, {\it Geophys.~Astrophys.~Fluid Dynamics}
{\bf 36}, 53--83 (1986).

\mi Galloway D.J. and Frisch U. A note on the stability of a
family of space-periodic Beltrami flows, {\it J. Fluid Mech.} {\bf 180},
557--564 (1987).

\mi Glatzmaier G.A. and Roberts P.H. A three-dimensional self-consistent
computer simulation of a geomagnetic field reversal.
{\it Nature} {\bf 377}, 203--209 (1995).

\mi Glatzmaier G.A. and Roberts P.H. An anelastic geodynamo simulation
driven by compositional and thermal convection, {\it Physica D} {\bf 97},
81--94 (1996).

\mi Hide R., Skeldon A.C. and Acheson D.J. A study of two novel
self-exiting single-disc homopolar dynamos: theory,
{\it Proc. R. Soc. Lond.} A {\bf 452}, 1369--1395 (1996).

\mi Jacobs J.A. {\it Reversals of the Earth's Magnetic Field},
Cambridge University Press, 1994.

\mi Jones C.A. Convection driven geodynamo models,
{\it Phil. Trans. R. Soc. Lond.} A {\bf 358}, 873--897 (2000).

\mi Matthews, P.C., Dynamo action in simple convective flows,
{\it Proc. R. Soc.} {\bf 455}, 1829--1840 (1999).

\mi Melbourne, I., Proctor, M.R.E. and Rucklidge, A.M.,
A heteroclinic model of geodynamo reversals and excursions,
In: {\it Dynamo and Dynamics, a Mathematical Challenge}
(Eds. Chossat, P., Armbruster, D. and Oprea, I.), 363--370,
Kluwer Academic Publishers (2001).

\mi Merrill R.T., McElhinny M.W. and Stevenson D.J. Evidence for long-term
asymmetries in the Earth's magnetic field and possible implications for dynamo
theories, {\it Phys.~Earth Planet.~Int.} {\bf 20}, 75--82 (1979).

\mi Merrill R.T. and McFadden P.L. Secular variation and the origin of
geomagnetic field reversals, {\it J. Geophys. Res.} {\bf 93}, 11589--11597 (1986).

\mi Moroz I.M., Hide R. and Soward A.M. On self-exiting coupled Faraday
disc homopolar dynamos driving series motors,
{\it Physica D} {\bf 117}, 128--144 (1998).

\mi Podvigina O. and Pouquet A. On the non-linear stability
of the 1:1:1 ABC flow, {\it Physica D} {\bf 75}, 471--508 (1994).

\mi Podvigina O.M. {\it Spatially-periodic evolutionary and steady
solutions to the three-dimensional Navier-Stokes equation with the ABC-force.}
Institute of Mechanics, Lomonosov Moscow State University (1999).

\mi Roberts P.H. and Glatzmaier G.A. The geodynamo , past, present
and future, {\it Geophys.~Astrophys.~Fluid Dynamics}
{\bf 94}, 47--84 (2001).

\mi Roberts P.H. and Stix M $\alpha$-Effect Dynamos, by the Bullard-Gellman
Formalism, {\it Astron. and Astrophys.} {\bf 18}, 453--466, (1972).

\mi Sarson G.R. and Jones C.A. A convection driven geodynamo reversal
model, {\it Phys.~Earth Planet.~Int.} {\bf 111}, 3--20 (1999).

\mi Sarson G.R. Reversal models from dynamo calculations
{\it Phil. Trans. Soc. Lond.} A {\bf 358}, 921--942 (2000).

\mi Zeldovich Ya.B., Ruzmaikin A.A. and Sokoloff D.D. {\it Magnetic fields
in astrophysics}, Gordon and Breach, Science Publishers (1983).

\pagebreak
\begin{figure}

\centerline{\psfig{file=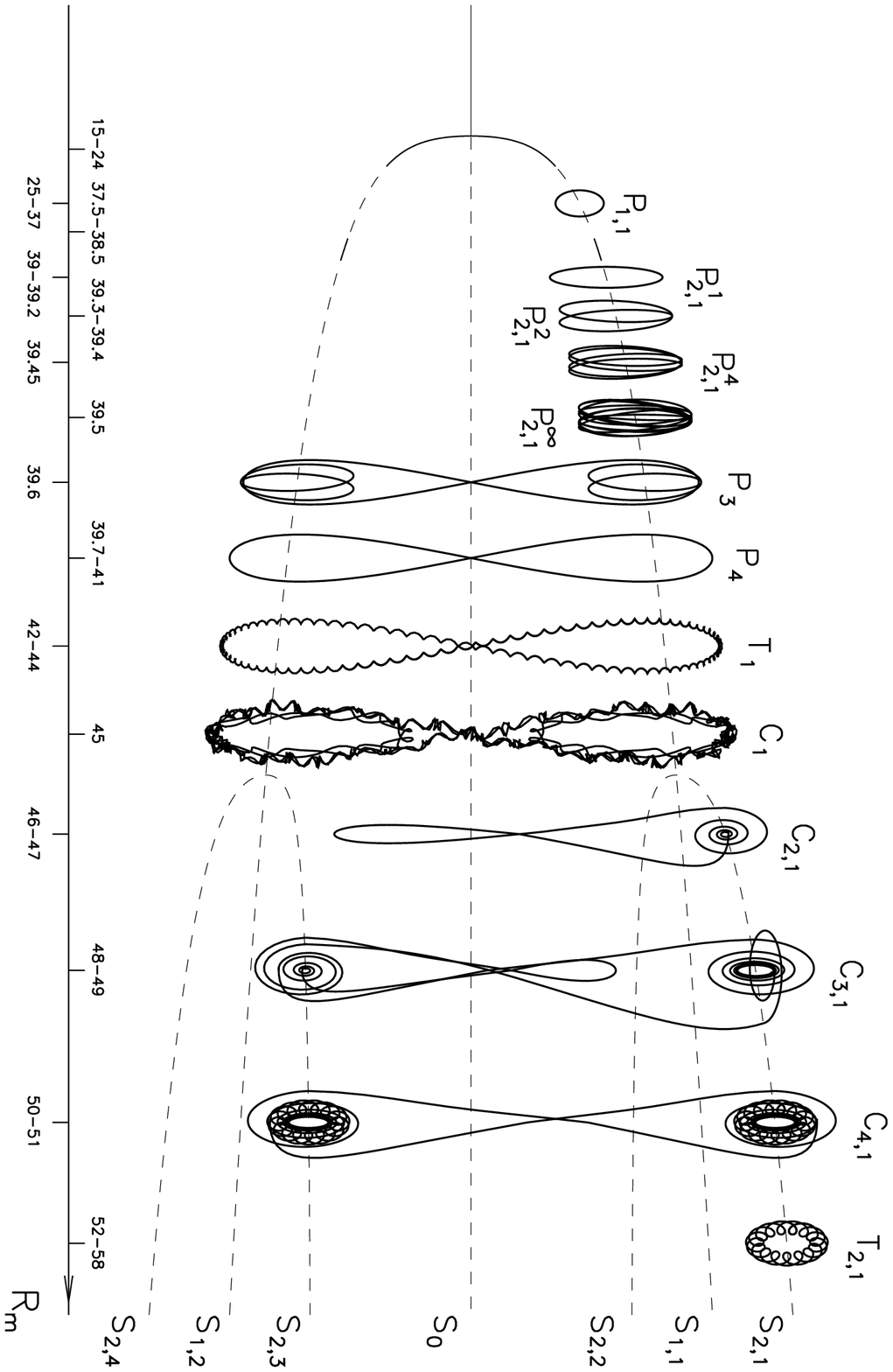,width=13cm,clip=}}
\vspace*{10mm}
\mi
Figure 1.
Bifurcation diagram of the detected MHD attractors for $R=4$ and $0<R_m\le 58$.
For each type of attractors the interval of $R_m$'s (horizontal axis,
non-uniform scale) is indicated, for which attractors of this type have been
detected in computations. Labeling of attractors is explained in Section 4
(see also Table 1). Only one representative symmetry-related attractors
is shown. The attractor $P_{2,1}^1$ for $R_m=39.6$ and attractors from
the second family are not shown. Stable steady states are represented
by solid lines, the unstable ones -- by dashed lines.
\end{figure}

\pagebreak
\begin{figure}

\centerline{\psfig{file=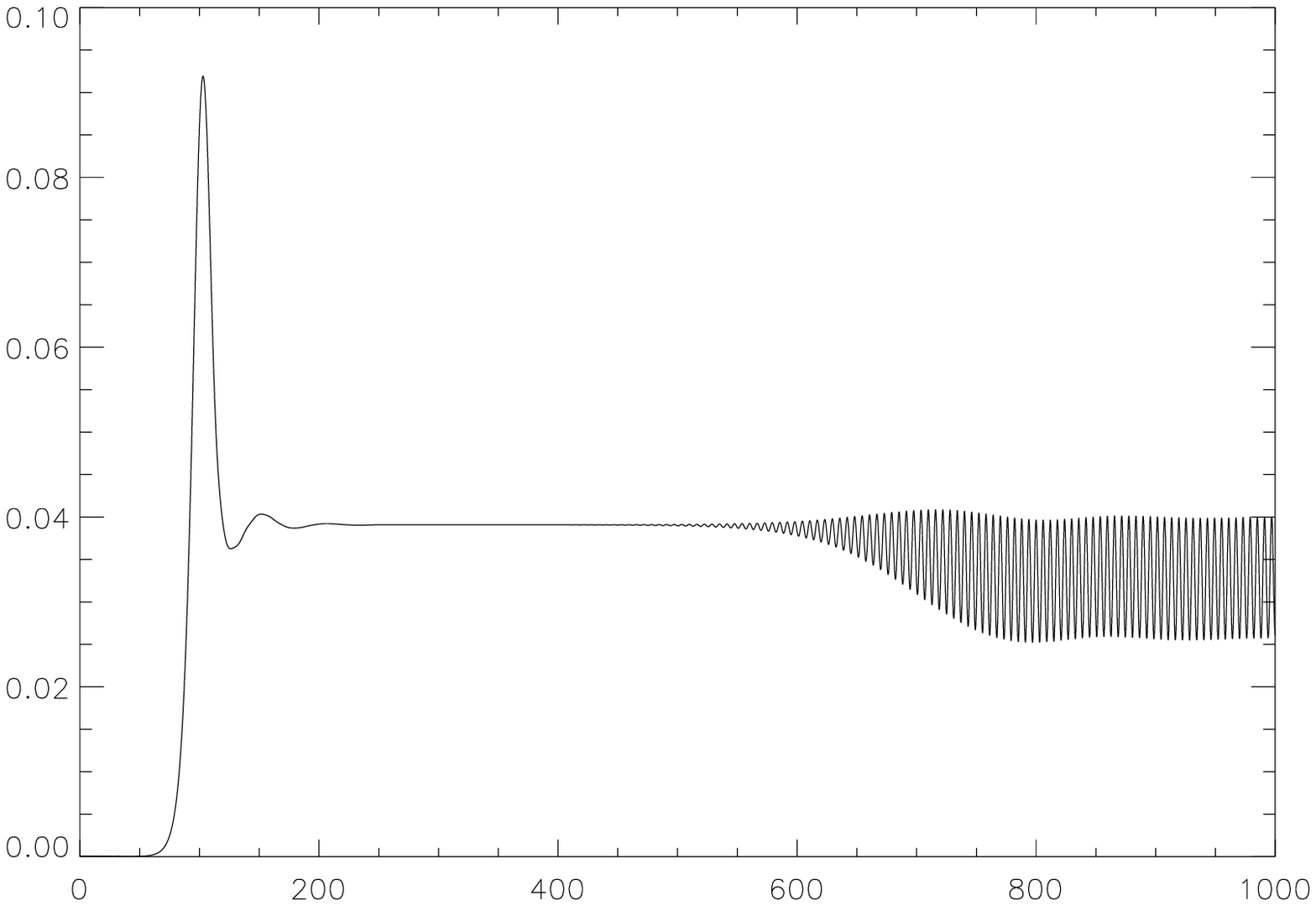,width=13cm,clip=}}
\vspace*{-5mm}
\mi
Figure 2a.
Magnetic energy (vertical axis) as a function of time (horizontal axis)
for $R=4$ and $R_m=30$.

\vspace*{10mm}

\centerline{\psfig{file=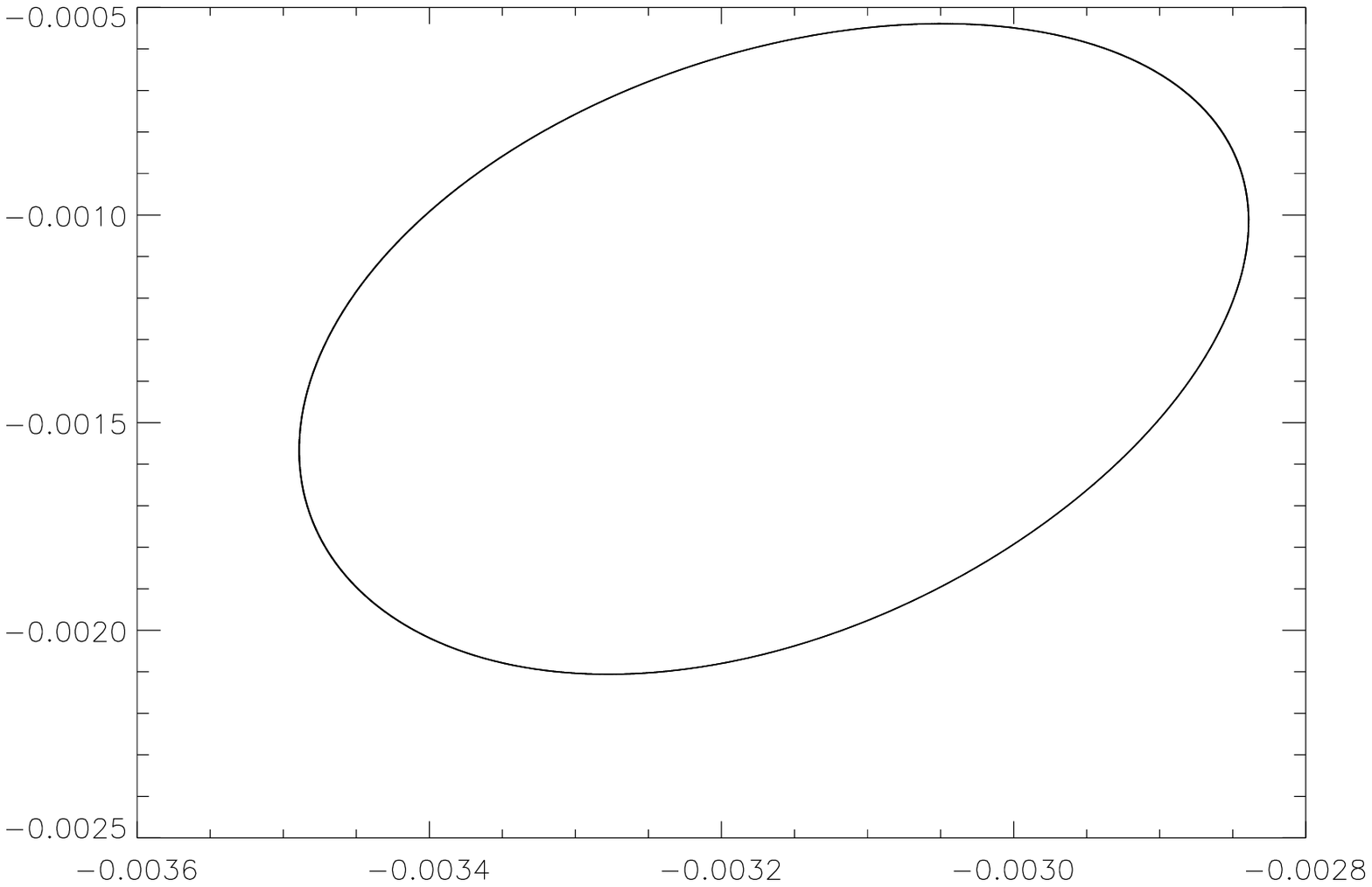,width=13cm,clip=}}
\vspace*{-5mm}
\mi
Figure 2b.
Projection of the trajectory in saturated regime (a periodic orbit $P_{1,i}$)
on the plane of Fourier coefficients Im$\,b^1_{0,1,2}$ (horizontal axis)
and Re$\,v^1_{0,1,2}$ (vertical axis) for $R=4$ and $R_m=30$
(same run as on Fig.~1a).
\end{figure}

\pagebreak
\begin{figure}

\centerline{\psfig{file=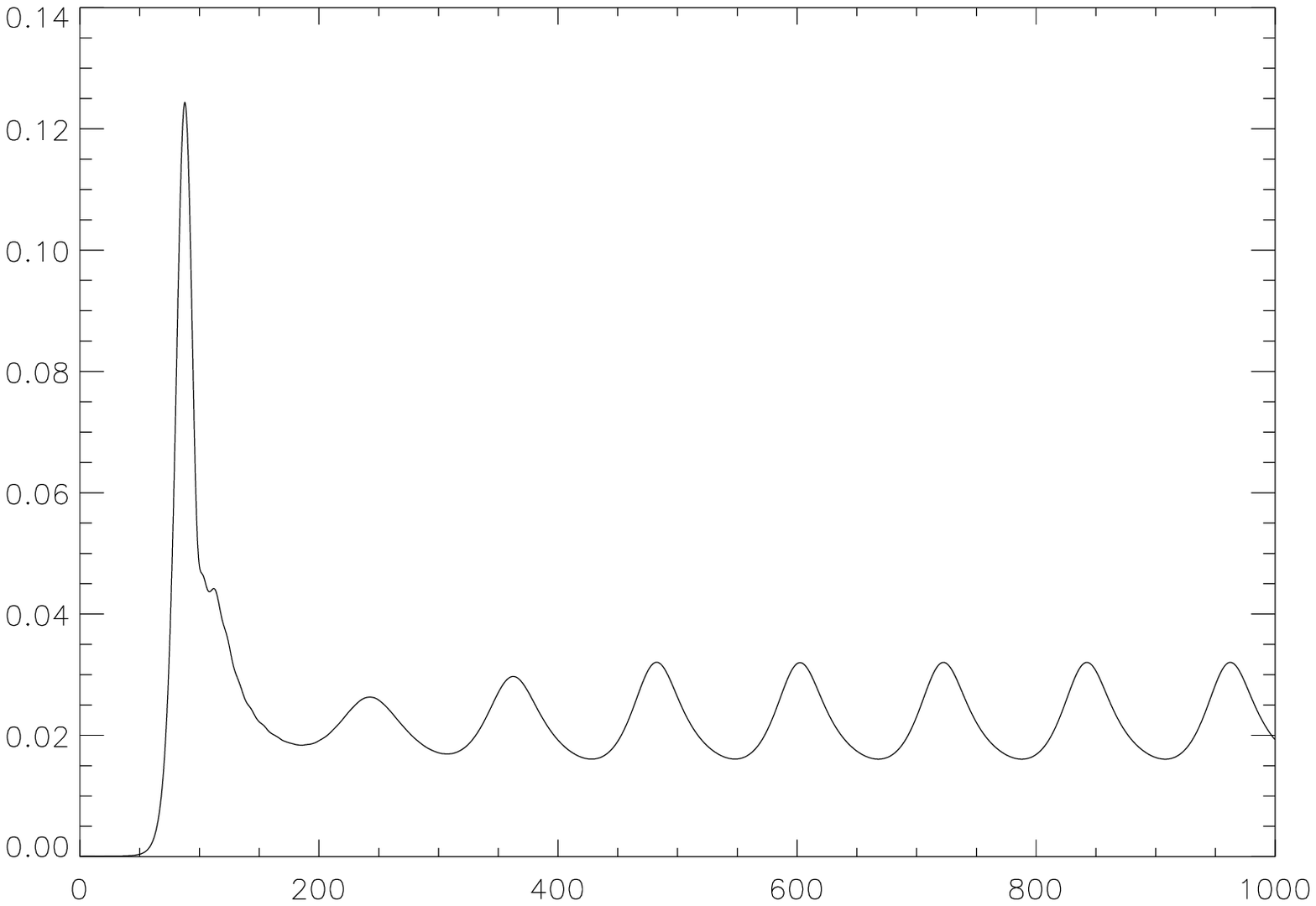,width=13cm,clip=}}
\vspace*{-5mm}
\mi
Figure 3a.
Magnetic energy (vertical axis) as a function of time (horizontal axis)
for $R=4$ and $R_m=39.2$.

\vspace*{10mm}
\centerline{\psfig{file=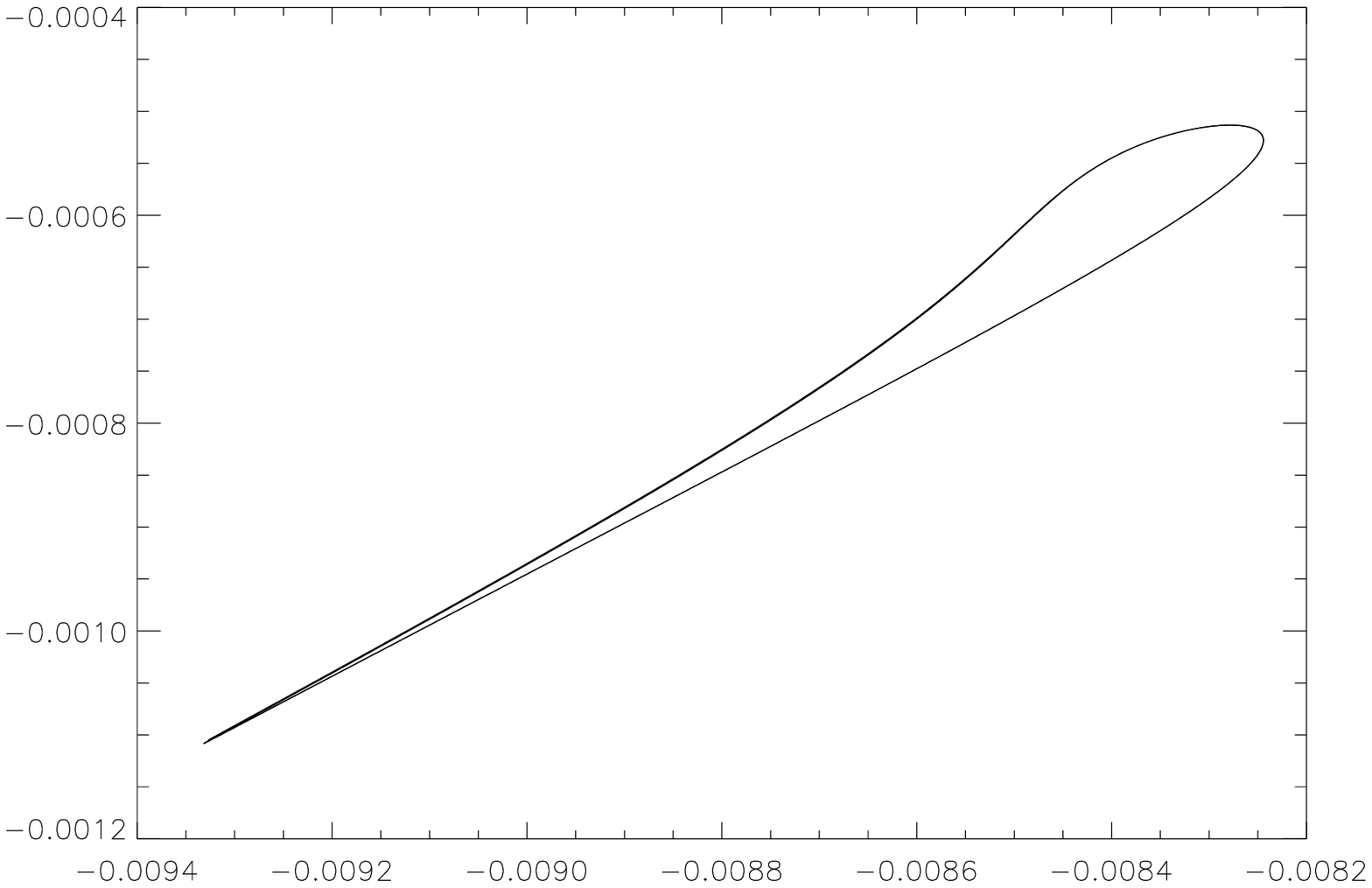,width=13cm,clip=}}
\vspace*{-5mm}
\mi
Figure 3b.
Projection of the trajectory in saturated regime (a periodic orbit $P_{2,i}^1$)
on the plane of Fourier coefficients Im$\,b^1_{0,1,2}$ (horizontal axis)
and Re$\,v^1_{0,1,2}$ (vertical axis) for $R=4$ and $R_m=39.2$
(same run as on Fig.~2a).
\end{figure}

\pagebreak
\begin{figure}

\centerline{\psfig{file=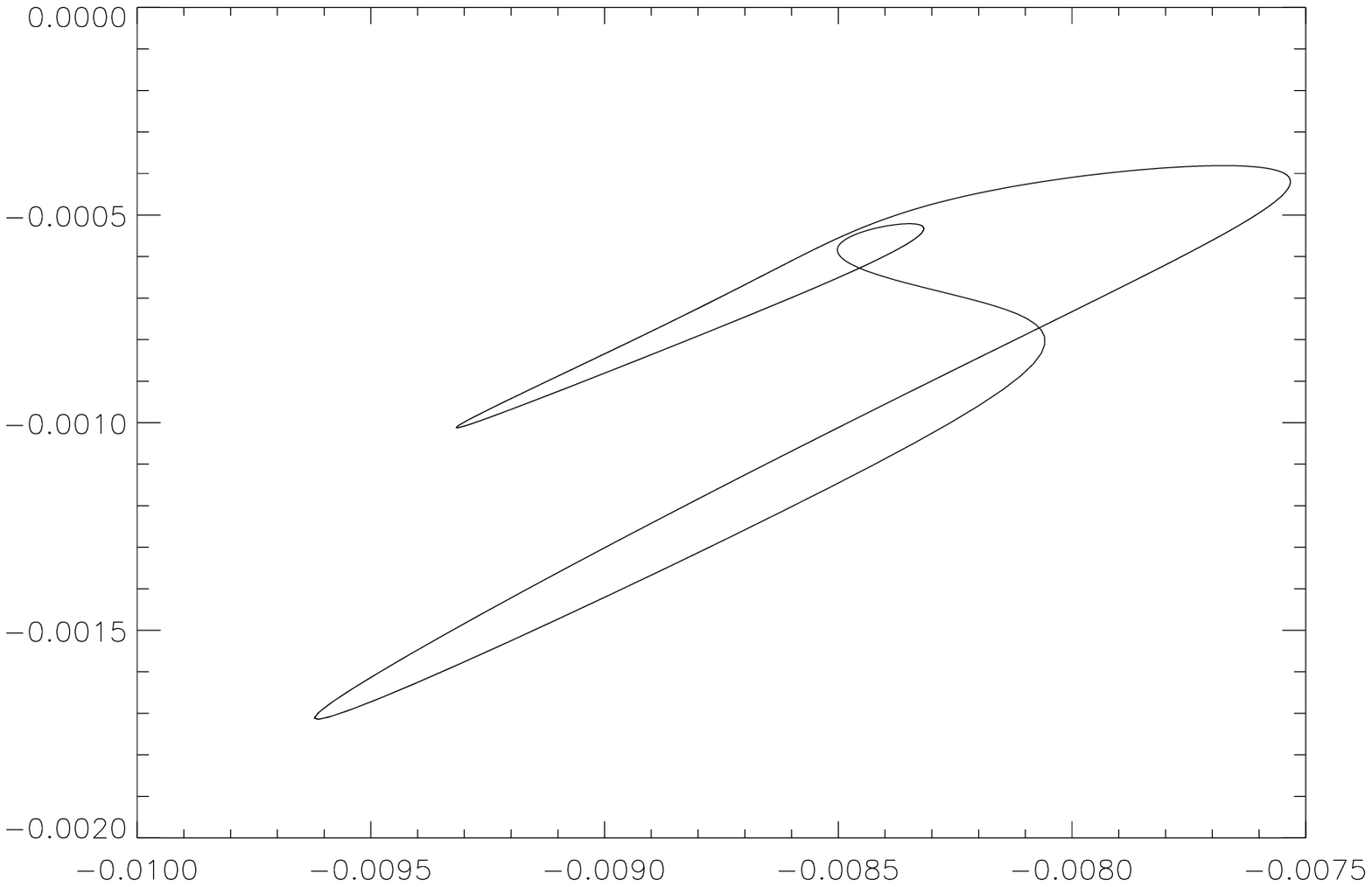,width=13cm,clip=}}
\vspace*{-5mm}
\mi
Figure 4a.
Projection of the trajectory in saturated regime (a periodic orbit $P_{2,i}^2$)
on the plane of Fourier coefficients Im$\,b^1_{0,1,2}$ (horizontal axis)
and Re$\,v^1_{0,1,2}$ (vertical axis) for $R=4$ and $R_m=39.4$.

\vspace*{10mm}
\centerline{\psfig{file=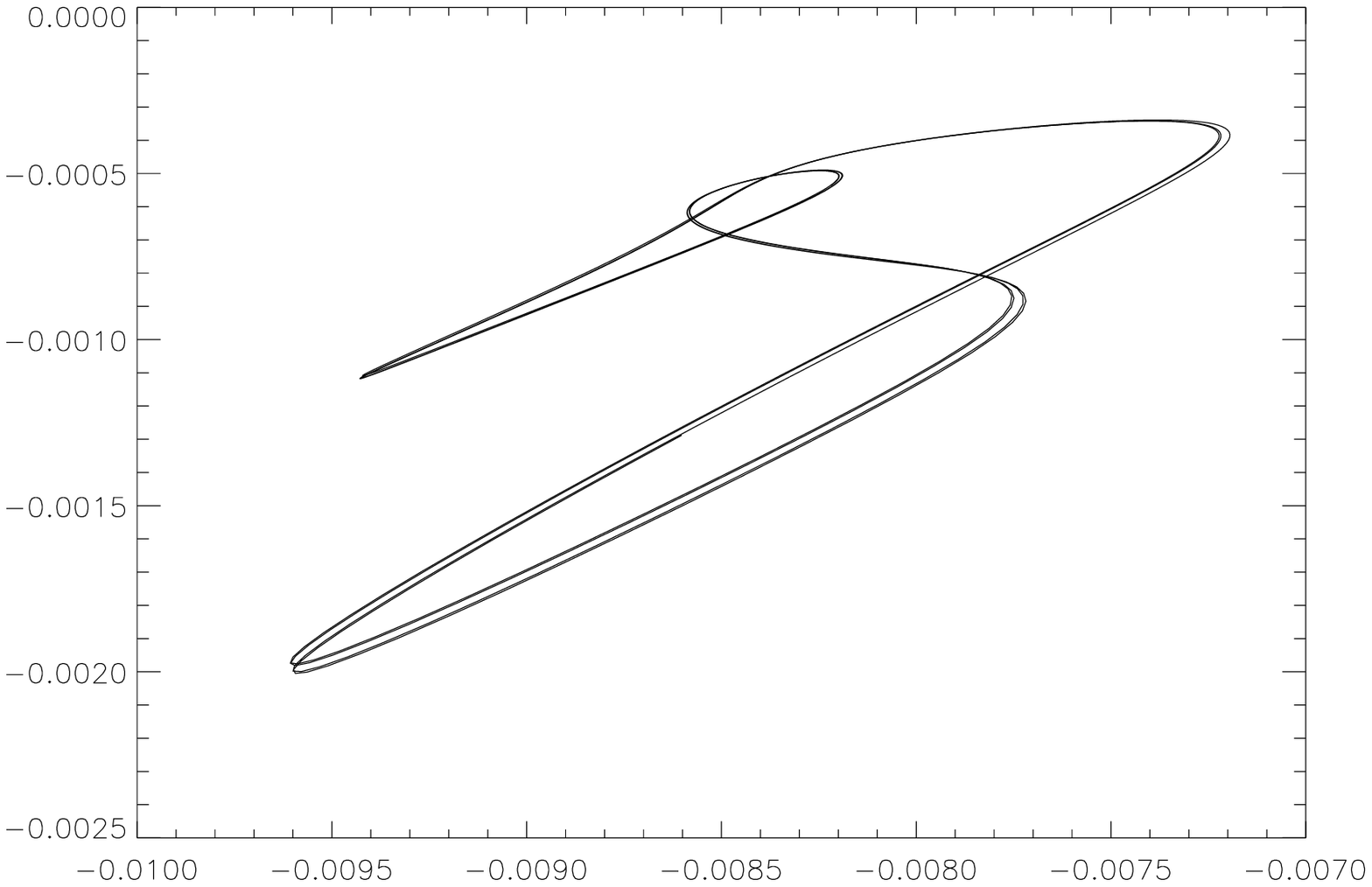,width=13cm,clip=}}
\vspace*{-5mm}
\mi
Figure 4b.
Projection of the trajectory in saturated regime (a periodic orbit $P_{2,i}^4$)
on the plane of Fourier coefficients Im$\,b^1_{0,1,2}$ (horizontal axis)
and Re$\,v^1_{0,1,2}$ (vertical axis) for $R=4$ and $R_m=39.45$.
\end{figure}

\pagebreak
\begin{figure}

\centerline{\psfig{file=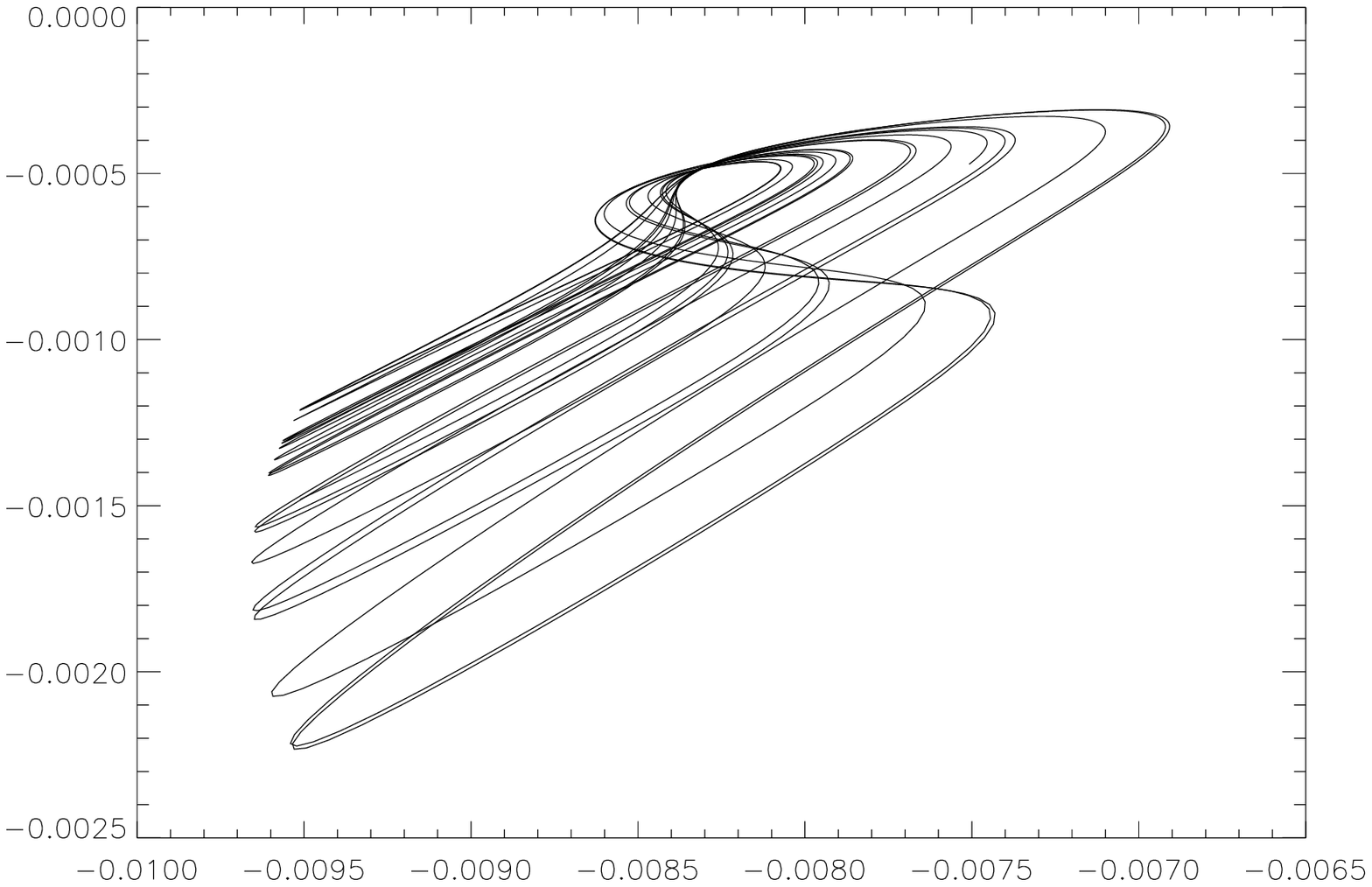,width=13cm,clip=}}
\vspace*{-5mm}
\mi
Figure 4c.
Projection of the trajectory in the phase space in saturated regime
(a periodic orbit $P_{2,i}^{\infty}$)
on the plane of Fourier coefficients Im$\,b^1_{0,1,2}$ (horizontal axis)
and Re$\,v^1_{0,1,2}$ (vertical axis) for $R=4$ and $R_m=39.5$.
\end{figure}

\pagebreak
\begin{figure}

\centerline{\psfig{file=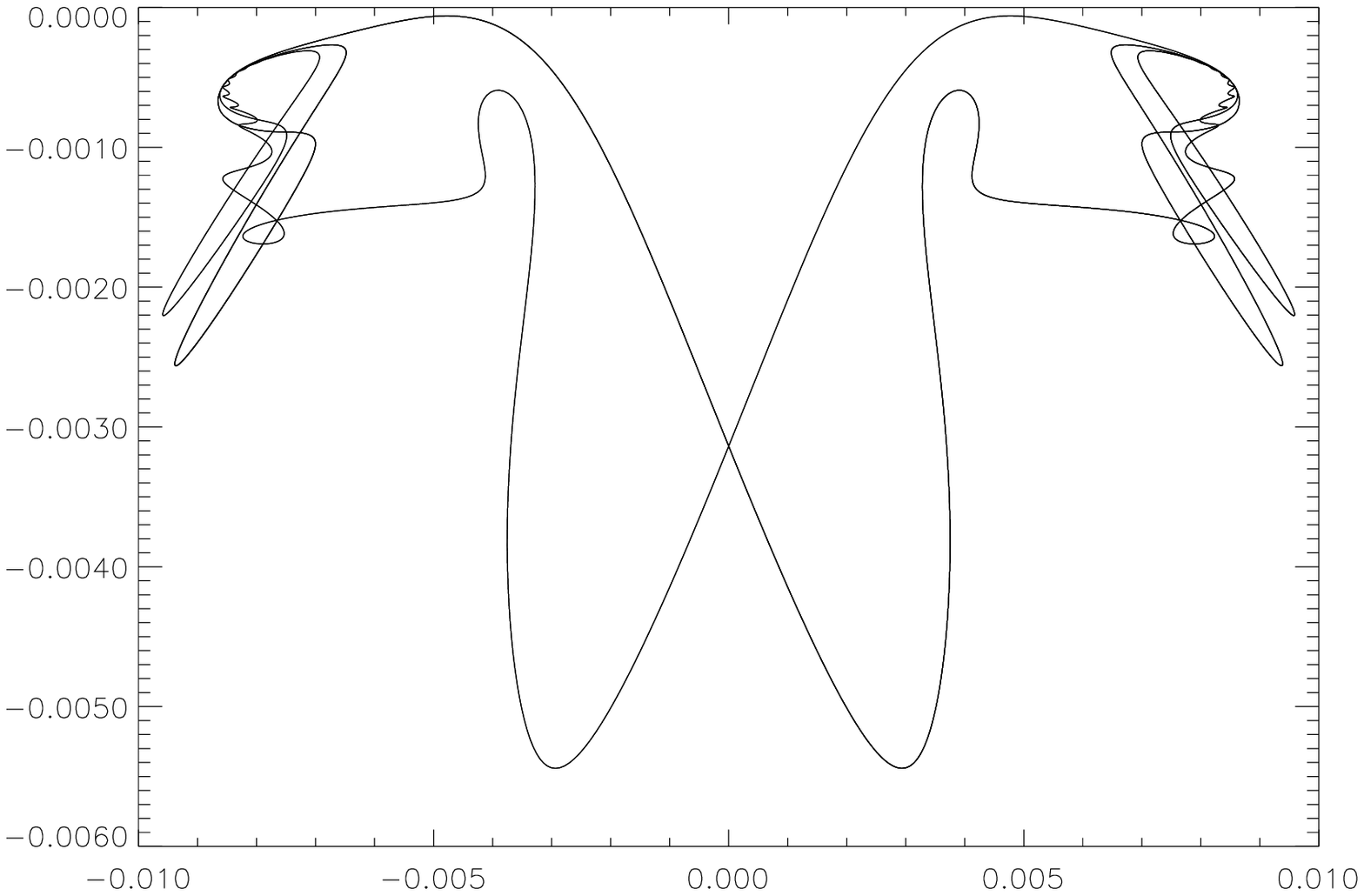,width=13cm,clip=}}
\vspace*{-5mm}
\mi
Figure 5a.
Projection of the trajectory in saturated regime (periodic orbit~$P_3$)
on the plane of Fourier coefficients Im$\,b^1_{0,1,2}$ (horizontal axis)
and Re$\,v^1_{0,1,2}$ (vertical axis) for $R=4$ and $R_m=39.6$.

\vspace*{10mm}
\centerline{\psfig{file=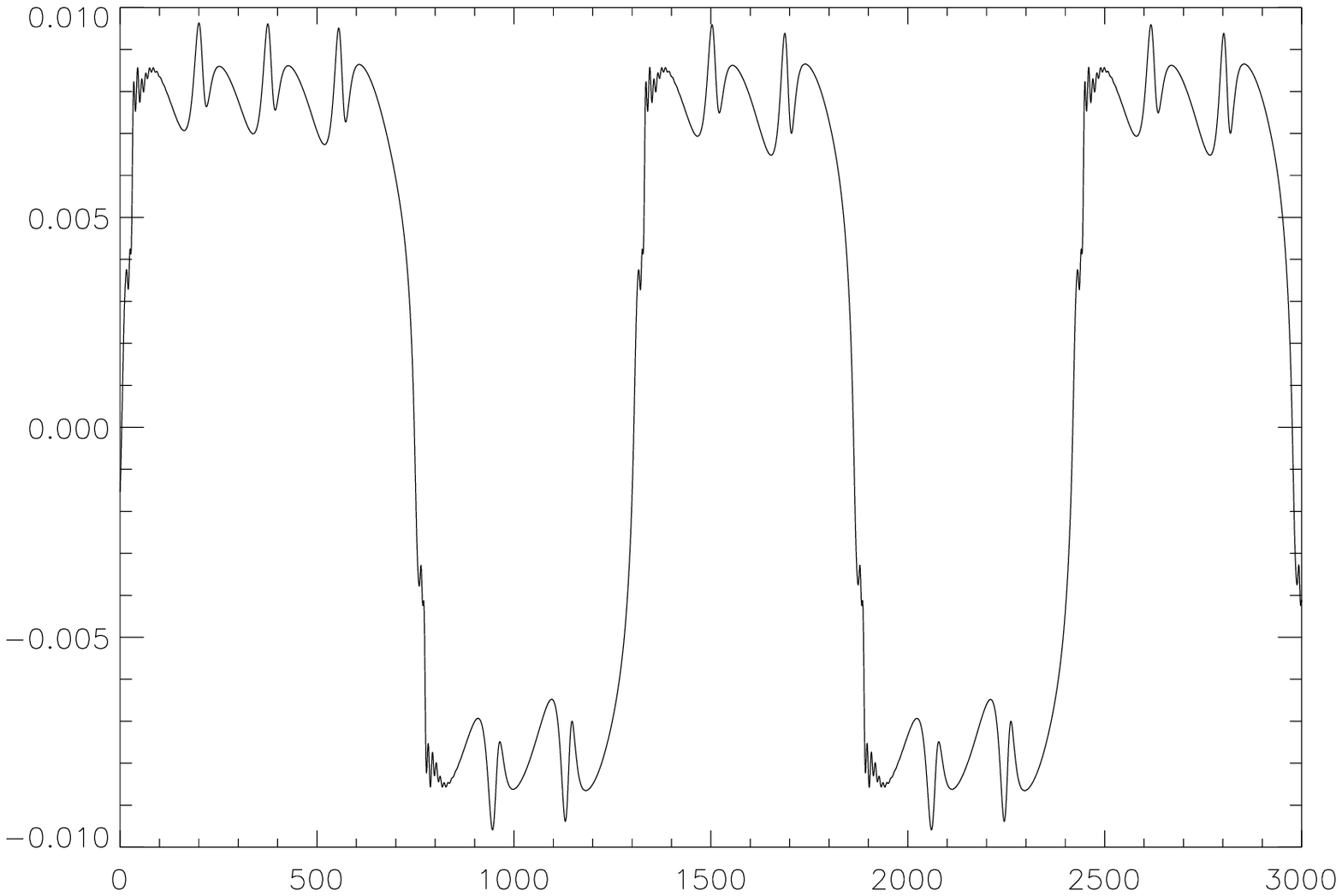,width=13cm,clip=}}
\vspace*{-5mm}
\mi
Figure 5b.
Fourier coefficient Im$\,b^1_{0,1,2}$ (vertical axis) as a function of time
(horizontal axis) for $R=4$ and $R_m=39.6$ (same run as on Fig.~4a).
\end{figure}

\pagebreak
\begin{figure}

\centerline{\psfig{file=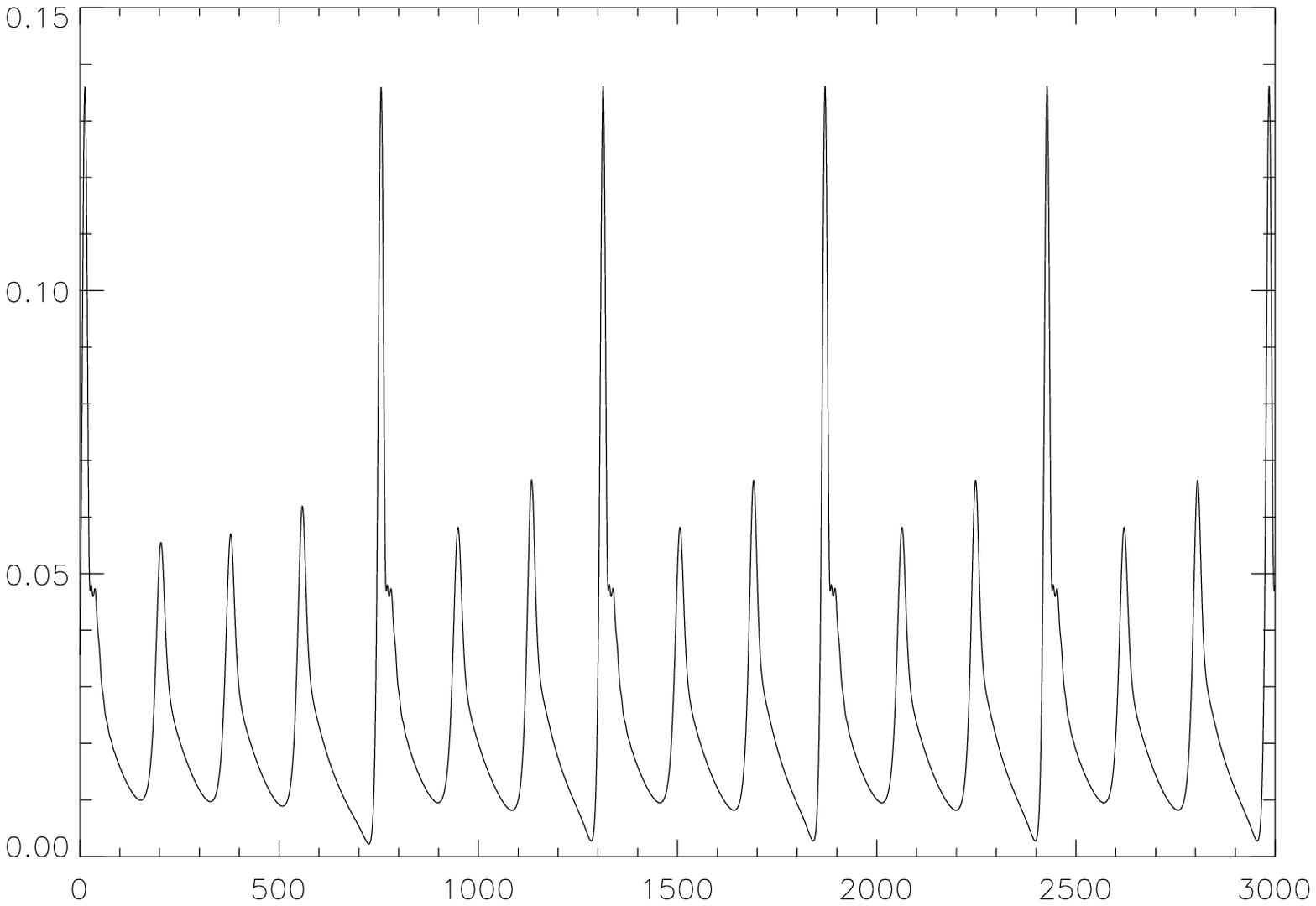,width=13cm,clip=}}
\vspace*{-5mm}
\mi
Figure 5c.
Magnetic energy (vertical axis) versus time (horizontal axis)
for $R=4$ and $R_m=39.6$ (same run as on Fig.~4a).

\vspace*{10mm}
\centerline{\psfig{file=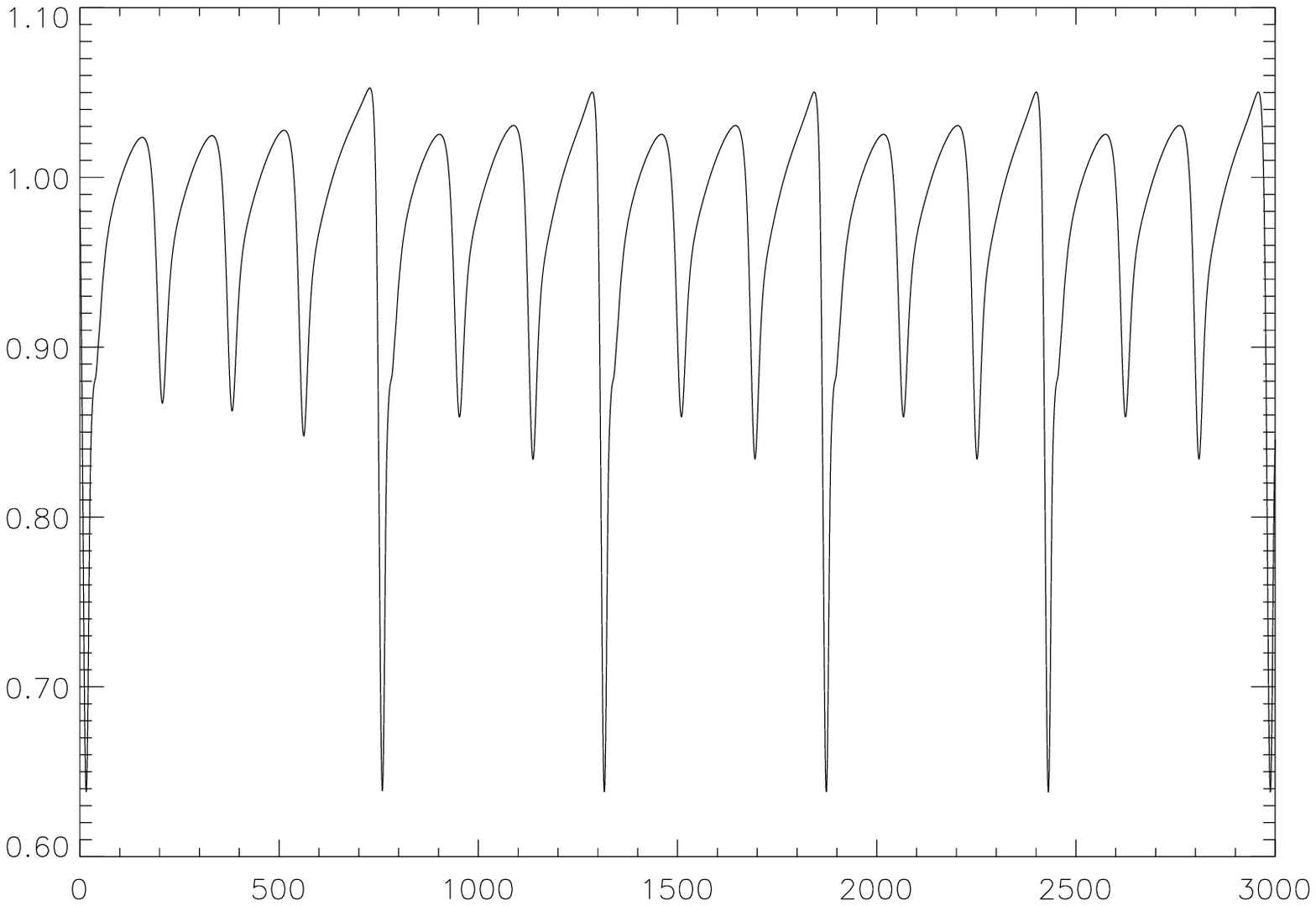,width=13cm,clip=}}
\vspace*{-5mm}
\mi
Figure 5d.
Kinetic energy (vertical axis) versus time (horizontal axis)
for $R=4$ and $R_m=39.6$ (same run as on Fig.~4a).
\end{figure}

\pagebreak
\begin{figure}

\centerline{\psfig{file=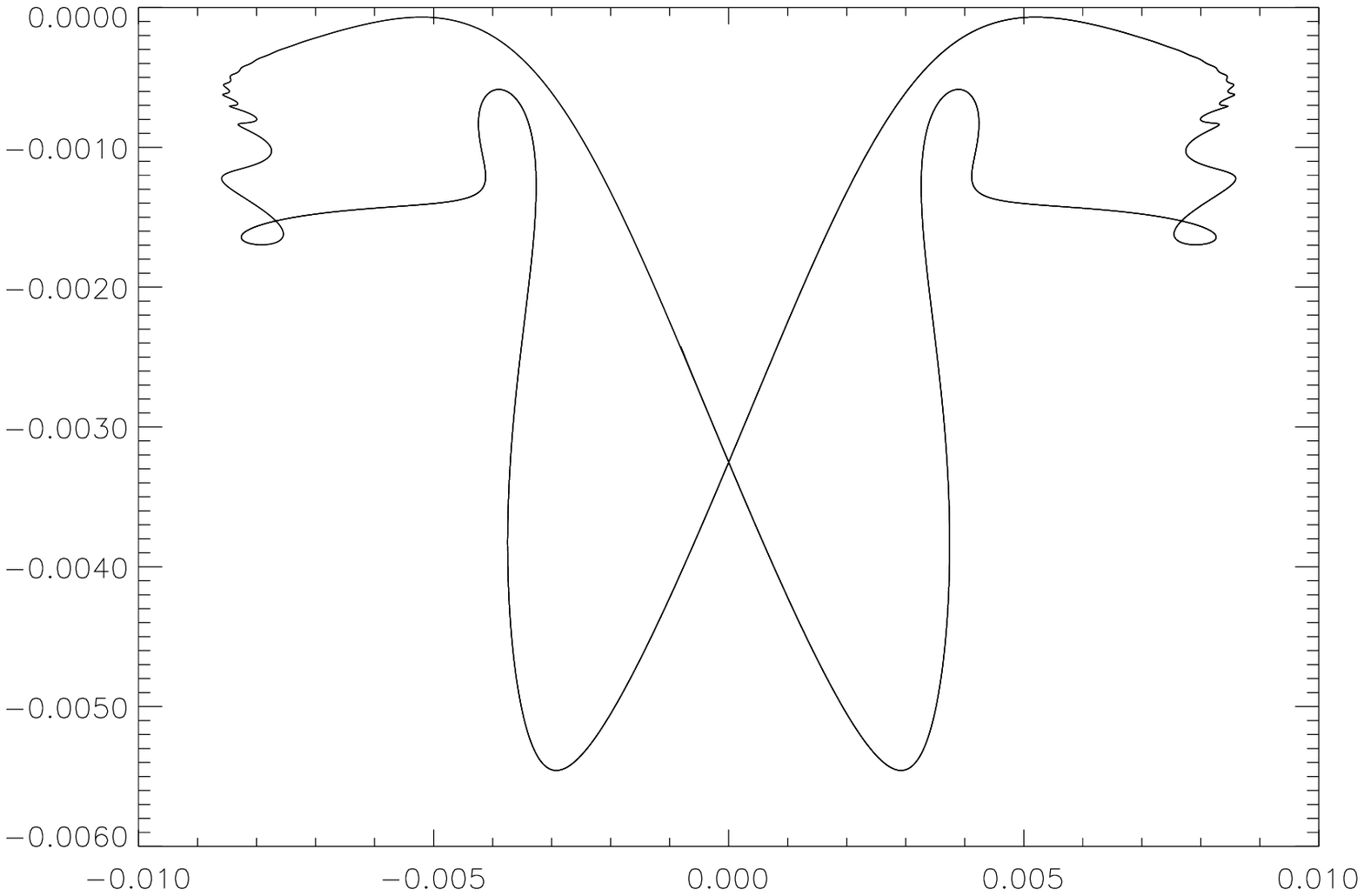,width=13cm,clip=}}
%\vspace*{-5mm}
\mi
Figure 6.
Projection of the trajectory in saturated regime (periodic orbit~$P_4$)
on the plane of Fourier coefficients Im$\,b^1_{0,1,2}$ (horizontal axis)
and Re$\,v^1_{0,1,2}$ (vertical axis) for $R=4$ and $R_m=39.7$.
\end{figure}

\pagebreak
\begin{figure}

\centerline{\psfig{file=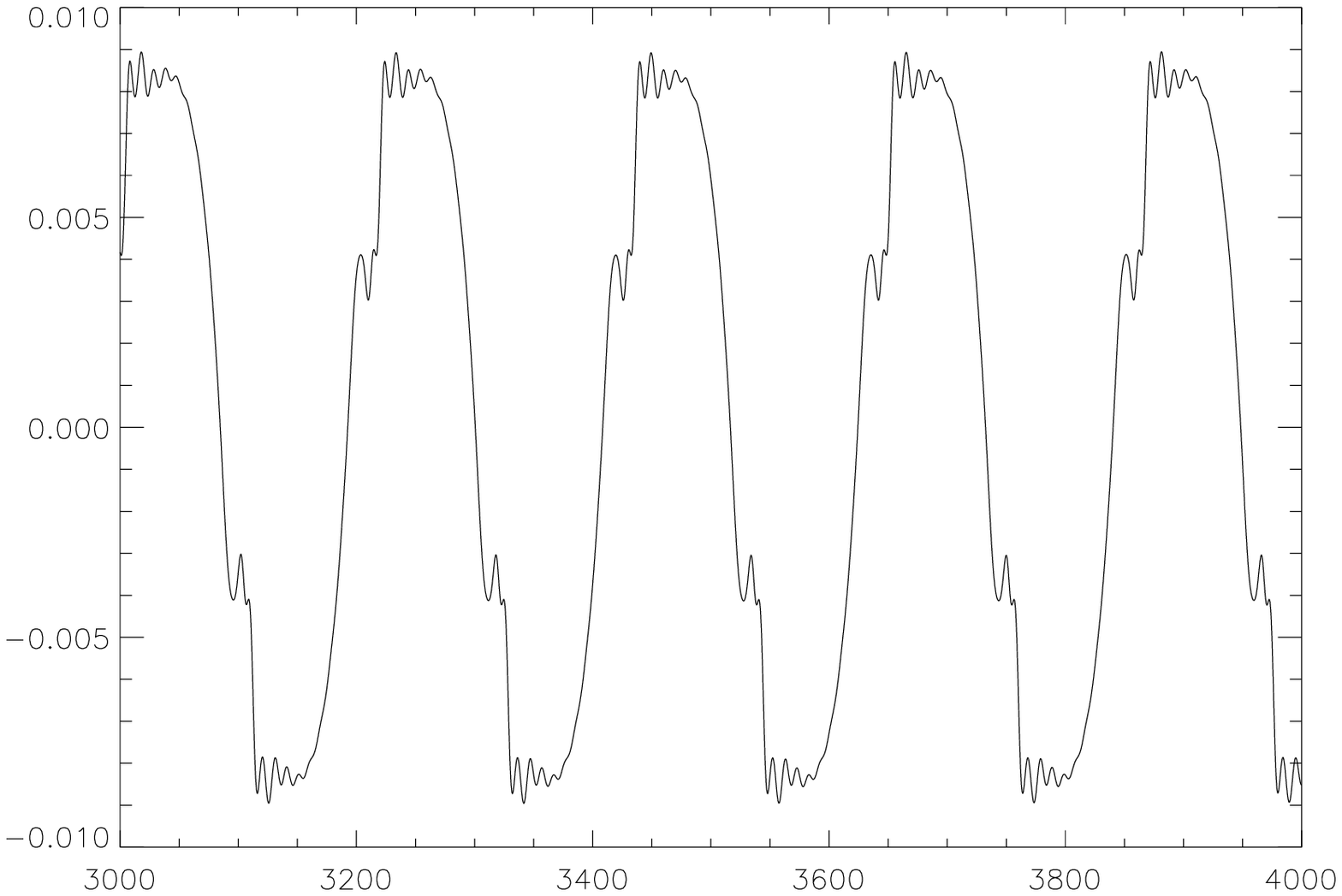,width=13cm,clip=}}
\vspace*{-5mm}
\mi
Figure 7a.
Fourier coefficient Im$\,b^1_{0,1,2}$ (vertical axis) as a function of time
(horizontal axis) for $R=4$ and $R_m=42$.

\vspace*{10mm}
\centerline{\psfig{file=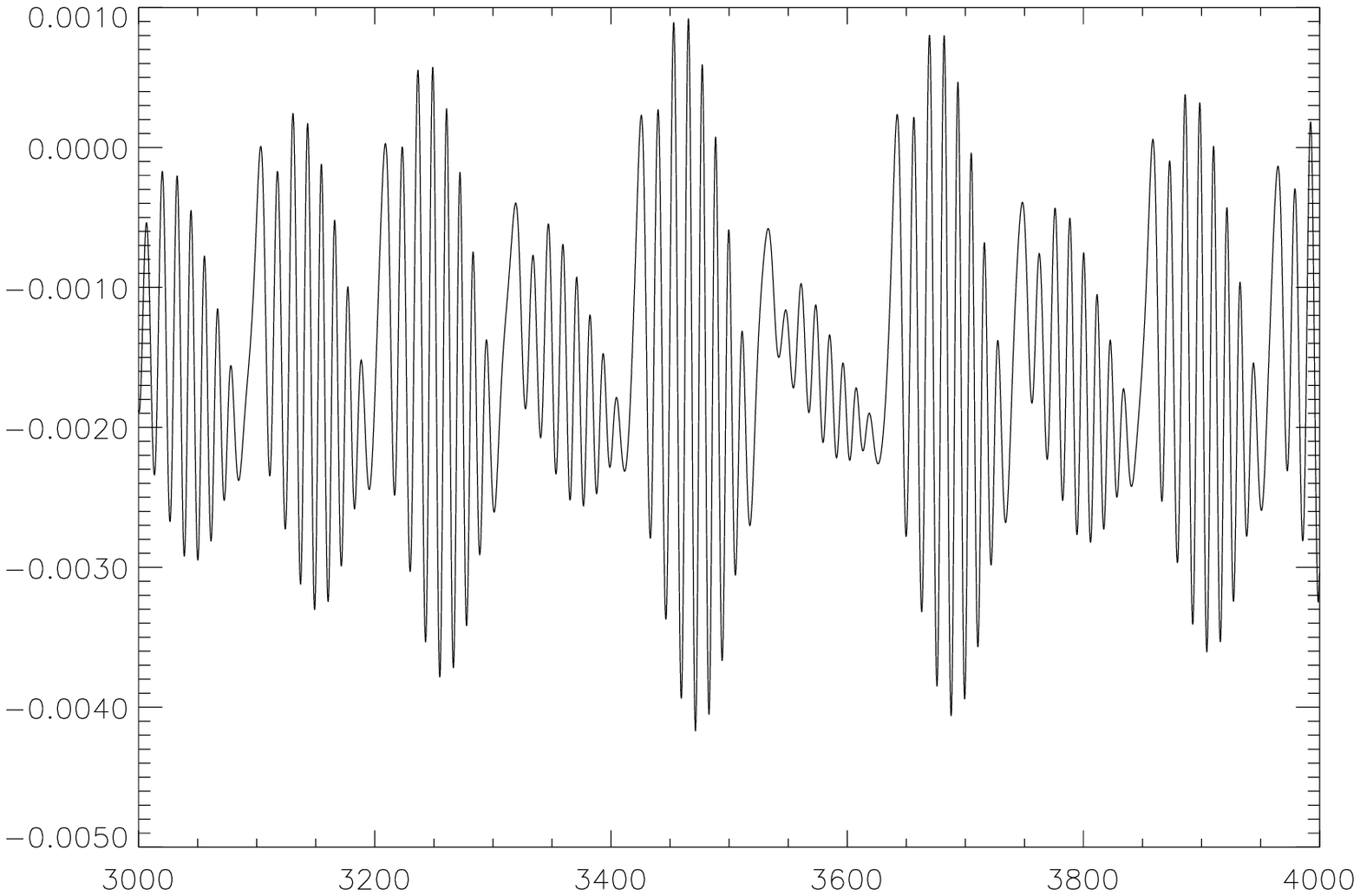,width=13cm,clip=}}
\vspace*{-5mm}
\mi
Figure 7b.
Fourier coefficient Re$\,b^3_{0,1,1}$ (vertical axis) as a function of time
(horizontal axis) for $R=4$ and $R_m=42$ (same run as on Fig.~6a).
\end{figure}

\pagebreak
\begin{figure}

\centerline{\psfig{file=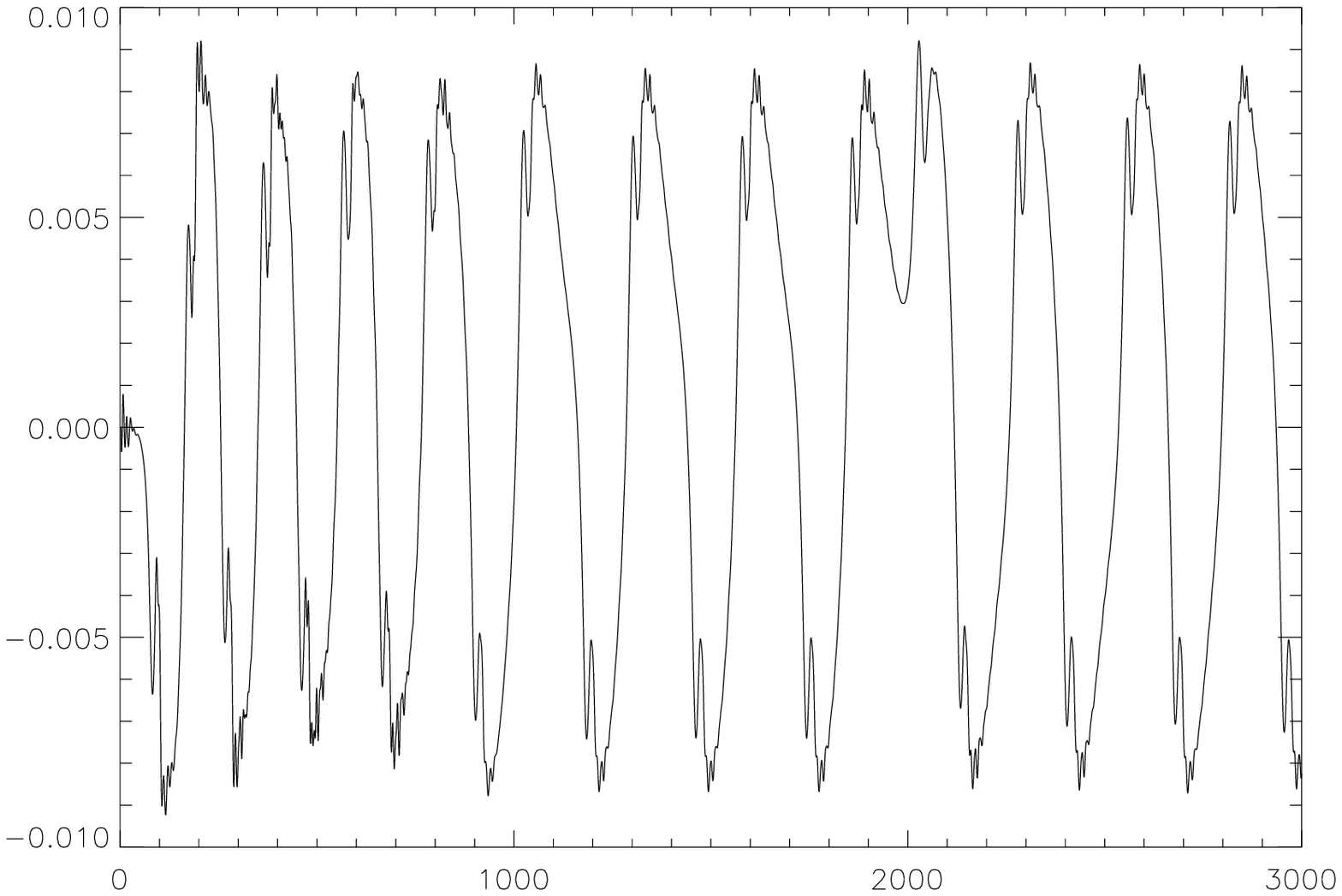,width=13cm,clip=}}
\vspace*{-5mm}
\mi
Figure 8a.
Fourier coefficient Im$\,b^1_{0,1,2}$ (vertical axis) as a function of time
(horizontal axis) for $R=4$ and $R_m=45$.

\vspace*{10mm}
\centerline{\psfig{file=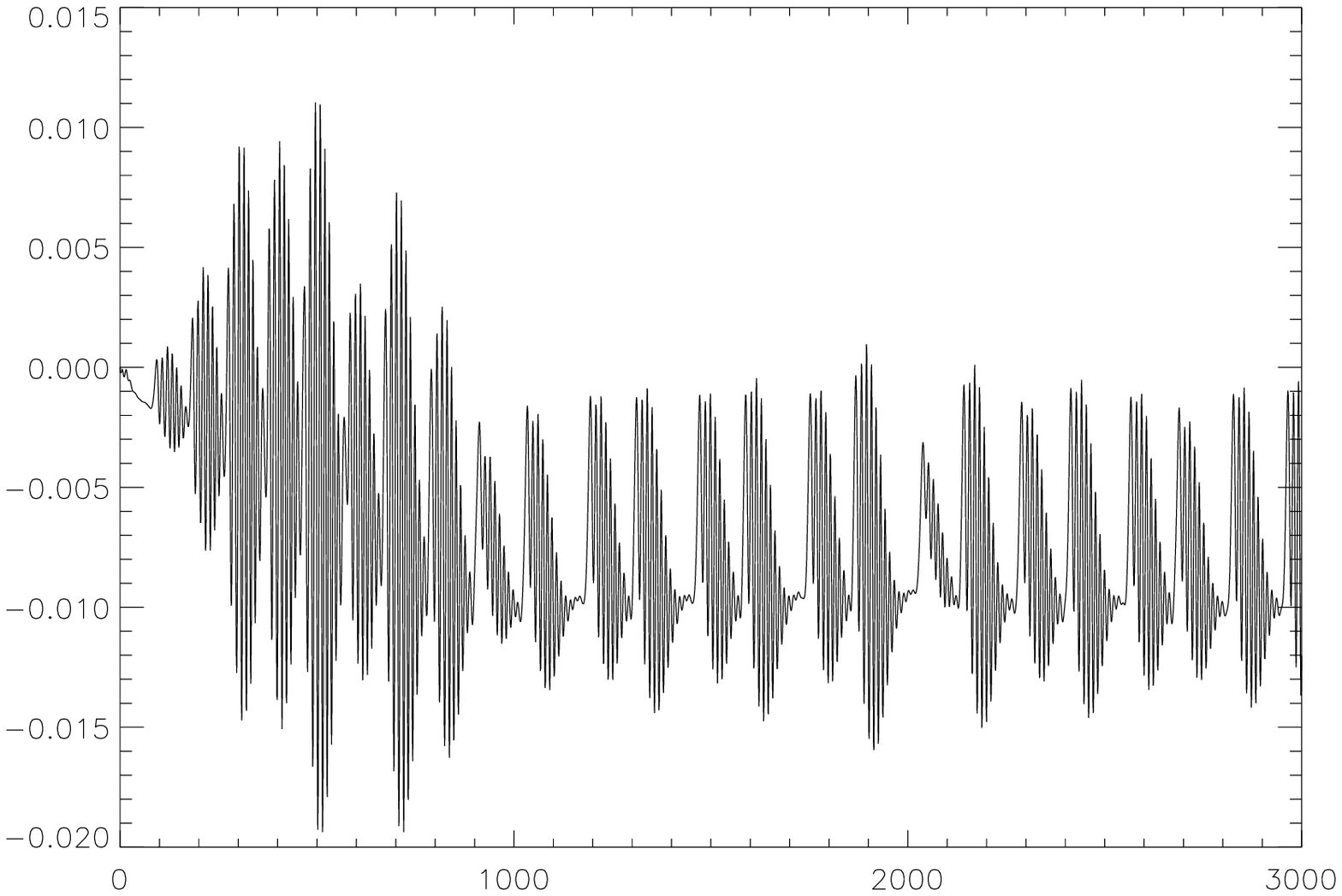,width=13cm,clip=}}
\vspace*{-5mm}
\mi
Figure 8b.
Fourier coefficient Re$\,b^3_{0,1,1}$ (vertical axis) as a function of time
(horizontal axis) for $R=4$ and $R_m=45$ (same run as on Fig.~7a).
\end{figure}

\pagebreak
\begin{figure}

\centerline{\psfig{file=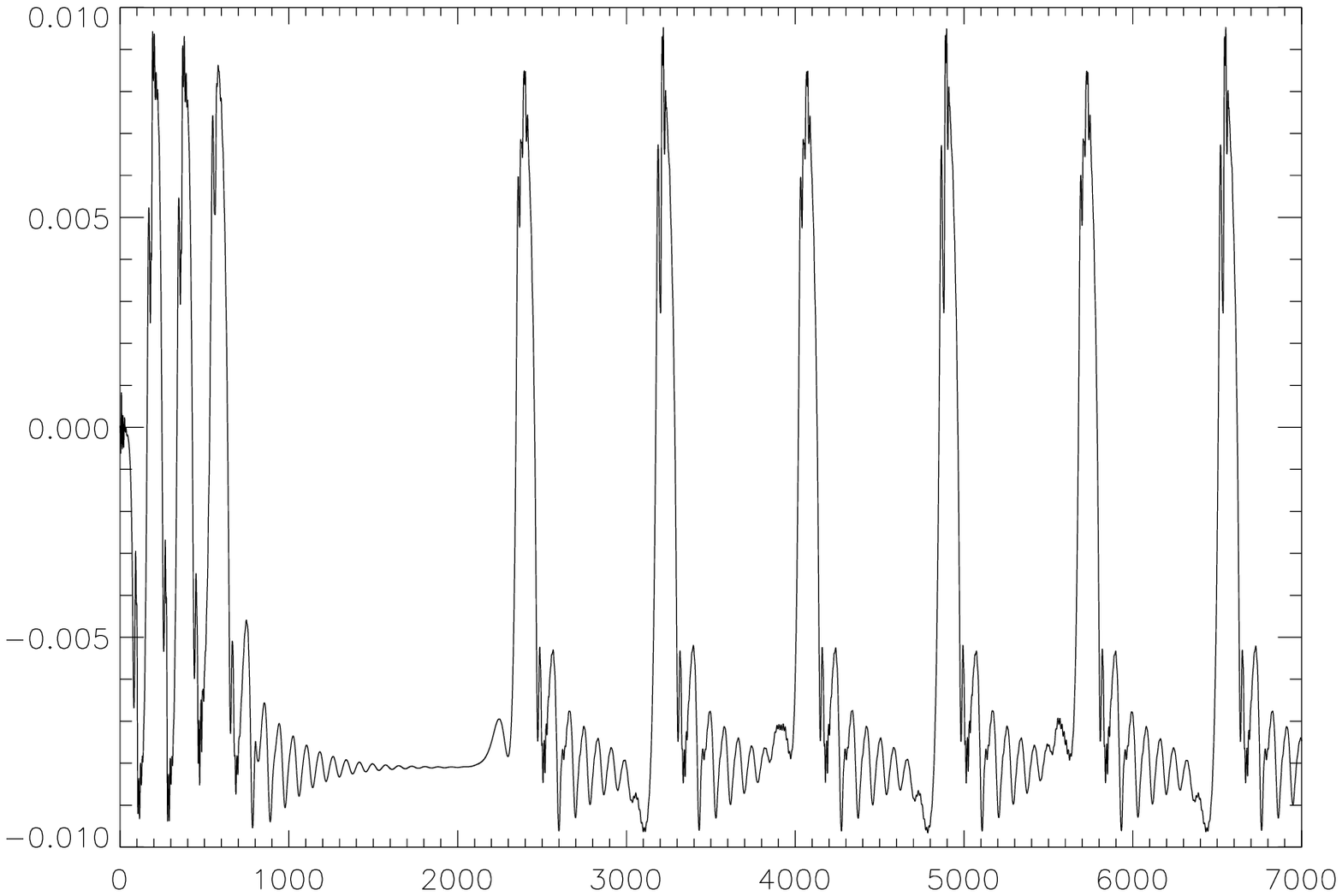,width=13cm,clip=}}
\vspace*{-5mm}
\mi
Figure 9a.
Fourier coefficient Im$\,b^1_{0,1,2}$ (vertical axis) as a function of time
(horizontal axis) for $R=4$ and $R_m=46$.

\vspace*{10mm}
\centerline{\psfig{file=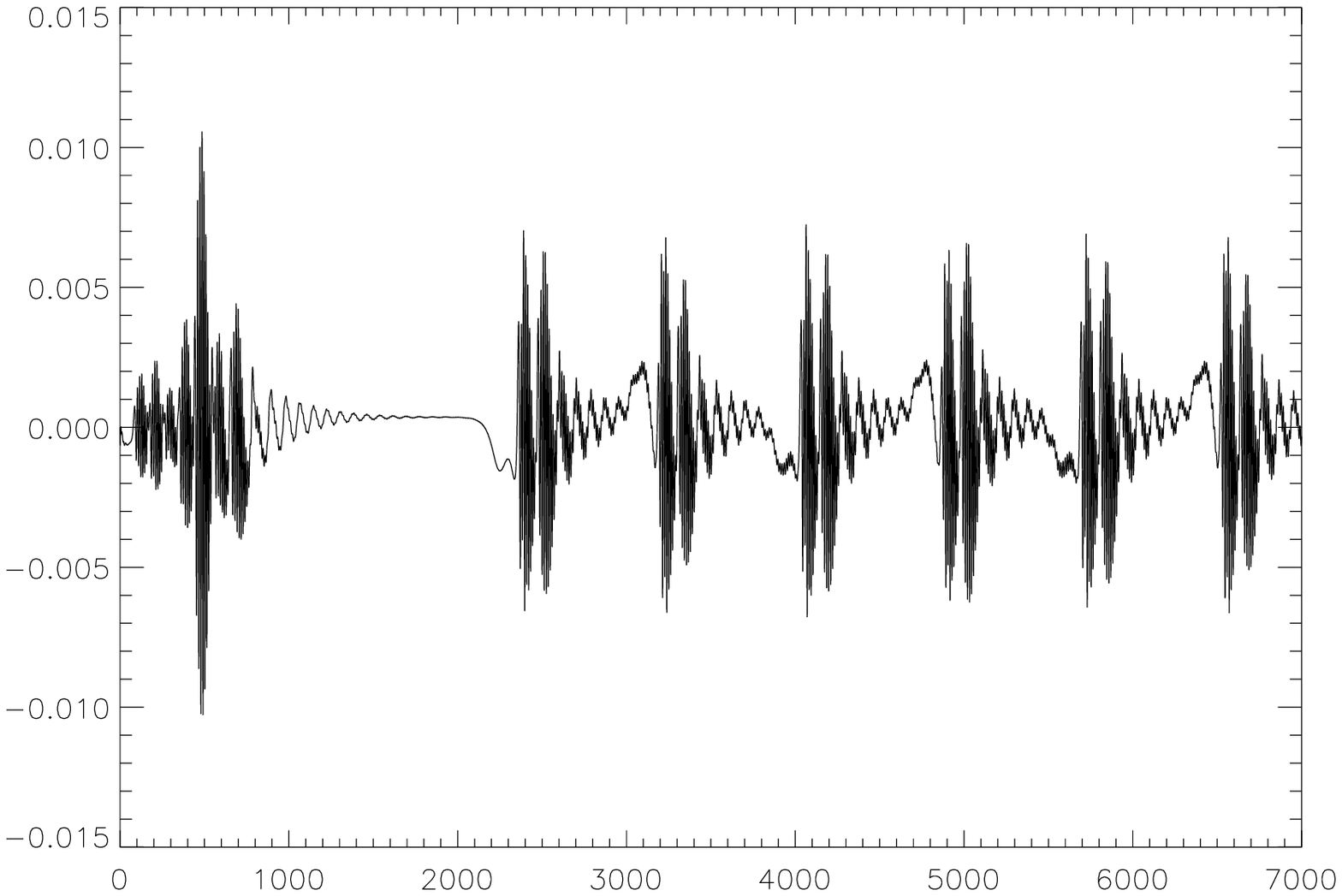,width=13cm,clip=}}
\vspace*{-5mm}
\mi
Figure 9b.
Fourier coefficient Re$\,b^3_{0,1,1}$ (vertical axis) as a function of time
(horizontal axis) for $R=4$ and $R_m=46$ (same run as on Fig.~8a).
\end{figure}

\pagebreak
\begin{figure}

\centerline{\psfig{file=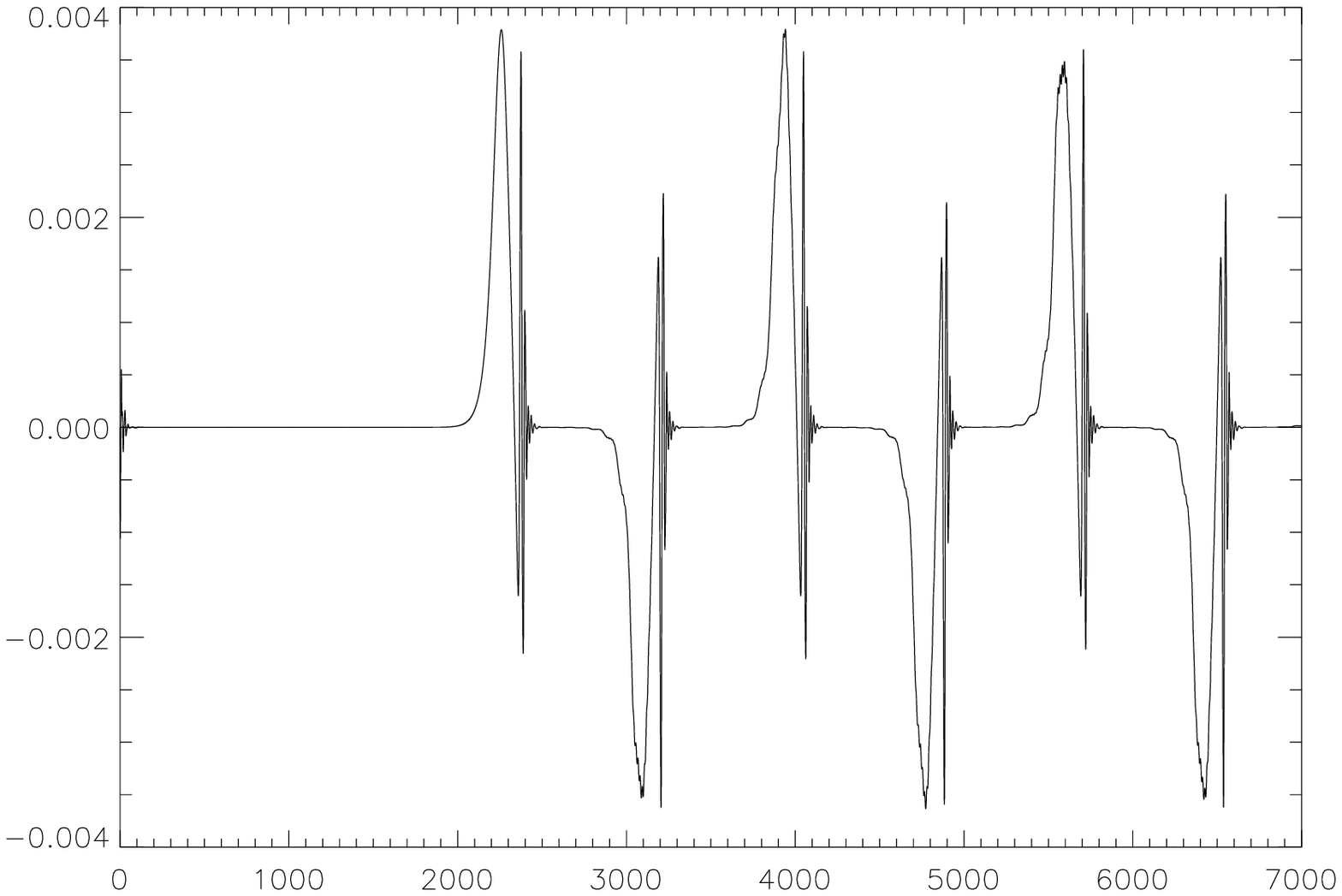,width=13cm,clip=}}
\vspace*{-5mm}
\mi
Figure 9c.
Fourier coefficient Re$\,b^3_{1,0,1}$ (vertical axis) as a function of time
(horizontal axis) for $R=4$ and $R_m=46$ (same run as on Fig.~8a).

\vspace*{10mm}
\centerline{\psfig{file=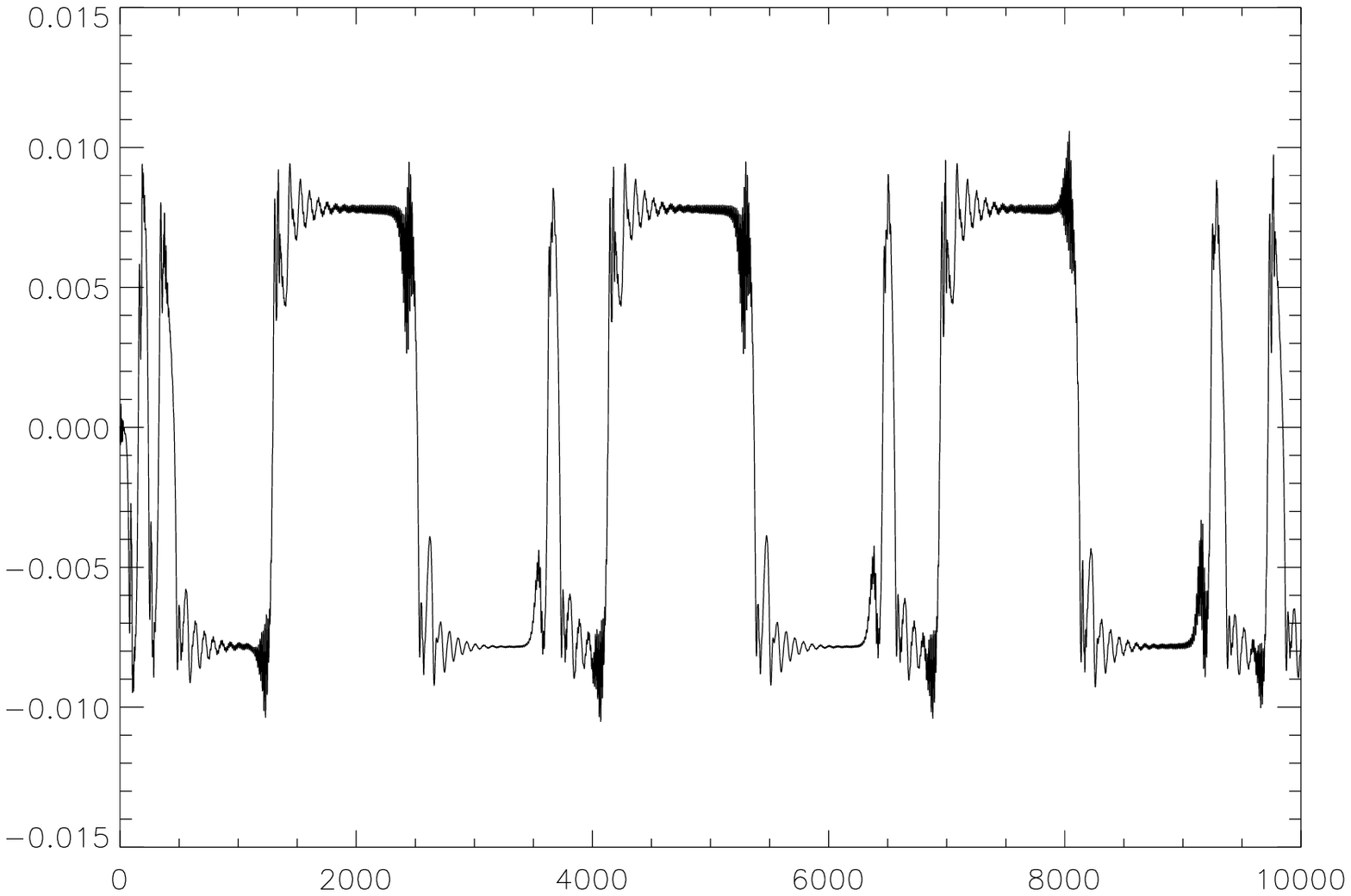,width=13cm,clip=}}
\vspace*{-5mm}
\mi
Figure 10.
Fourier coefficient Im$\,b^1_{0,1,2}$ (vertical axis) as a function of time
(horizontal axis) for $R=4$ and $R_m=48$.
\end{figure}

\pagebreak
\begin{figure}

\centerline{\psfig{file=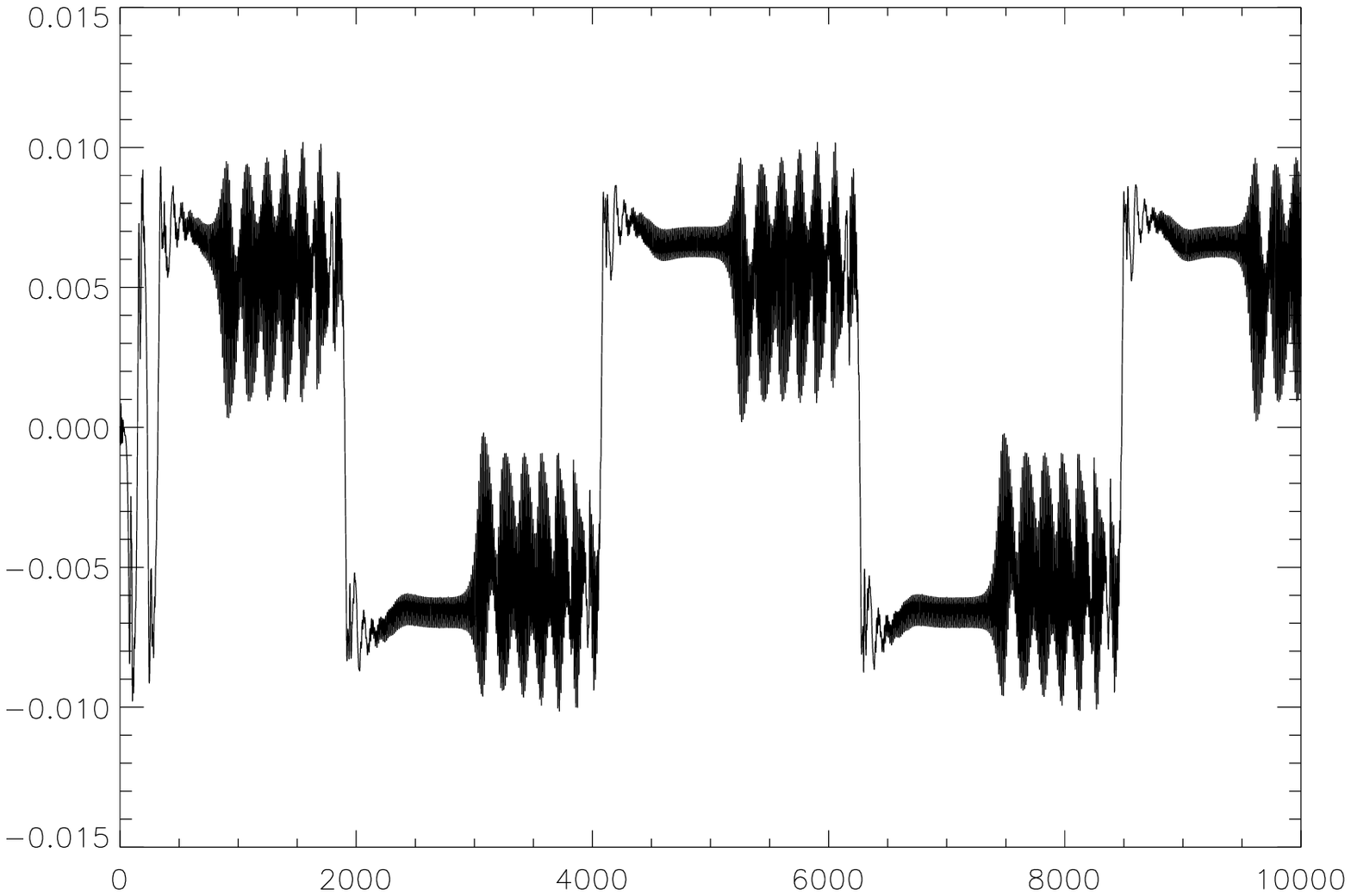,width=13cm,clip=}}
\vspace*{-5mm}
\mi
Figure 11a.
Fourier coefficients Im$\,b^1_{0,1,2}$ (vertical axis) as a function of time
(horizontal axis) for $R=4$ and $R_m=51$.

\vspace*{10mm}
\centerline{\psfig{file=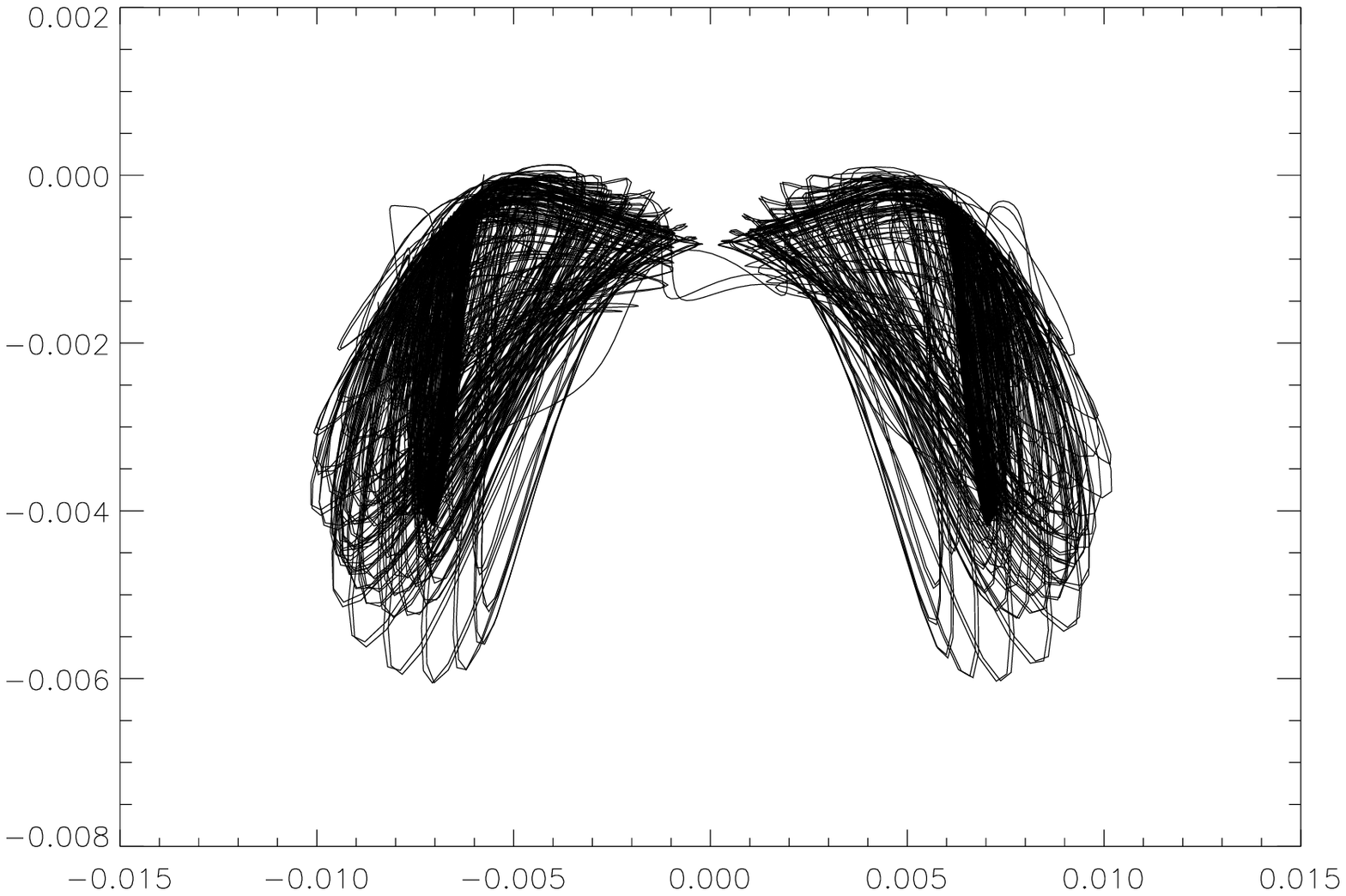,width=13cm,clip=}}
\vspace*{-5mm}
\mi
Figure 11b.
Projection of the trajectory in saturated regime
on the plane of Fourier coefficients Im$\,b^1_{0,1,2}$
(horizontal axis) and Re$\,v^1_{0,1,2}$ (vertical axis) for $R=4$ and
$R_m=51$ (same run as on Fig.~10a).
\end{figure}

\pagebreak
\begin{figure}

\centerline{\psfig{file=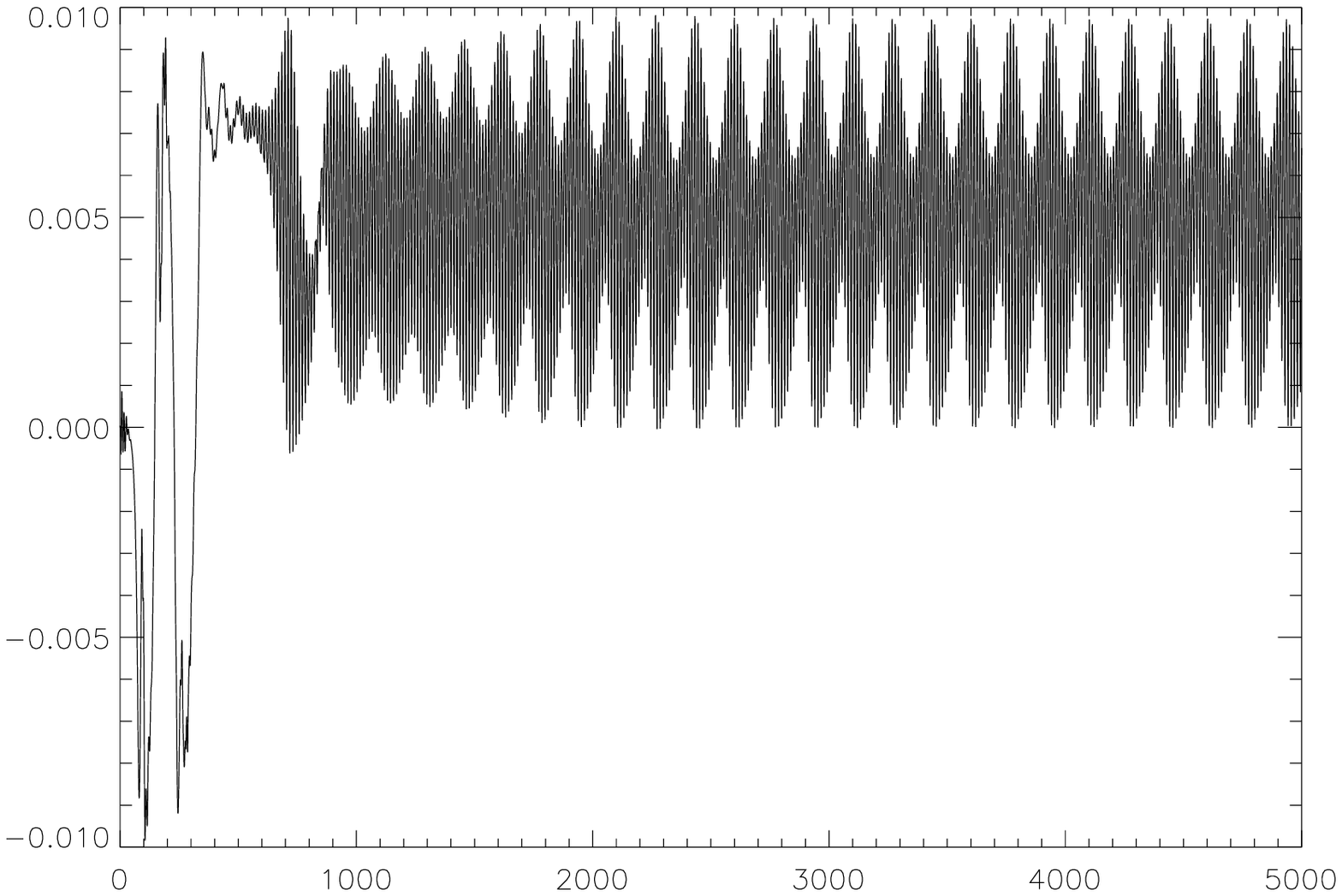,width=13cm,clip=}}
\vspace*{-5mm}
\mi
Figure 12a.
Fourier coefficient Im$\,b^1_{0,1,2}$ (vertical axis) as a function of time
(horizontal axis) for $R=4$ and $R_m=52$.

\vspace*{10mm}
\centerline{\psfig{file=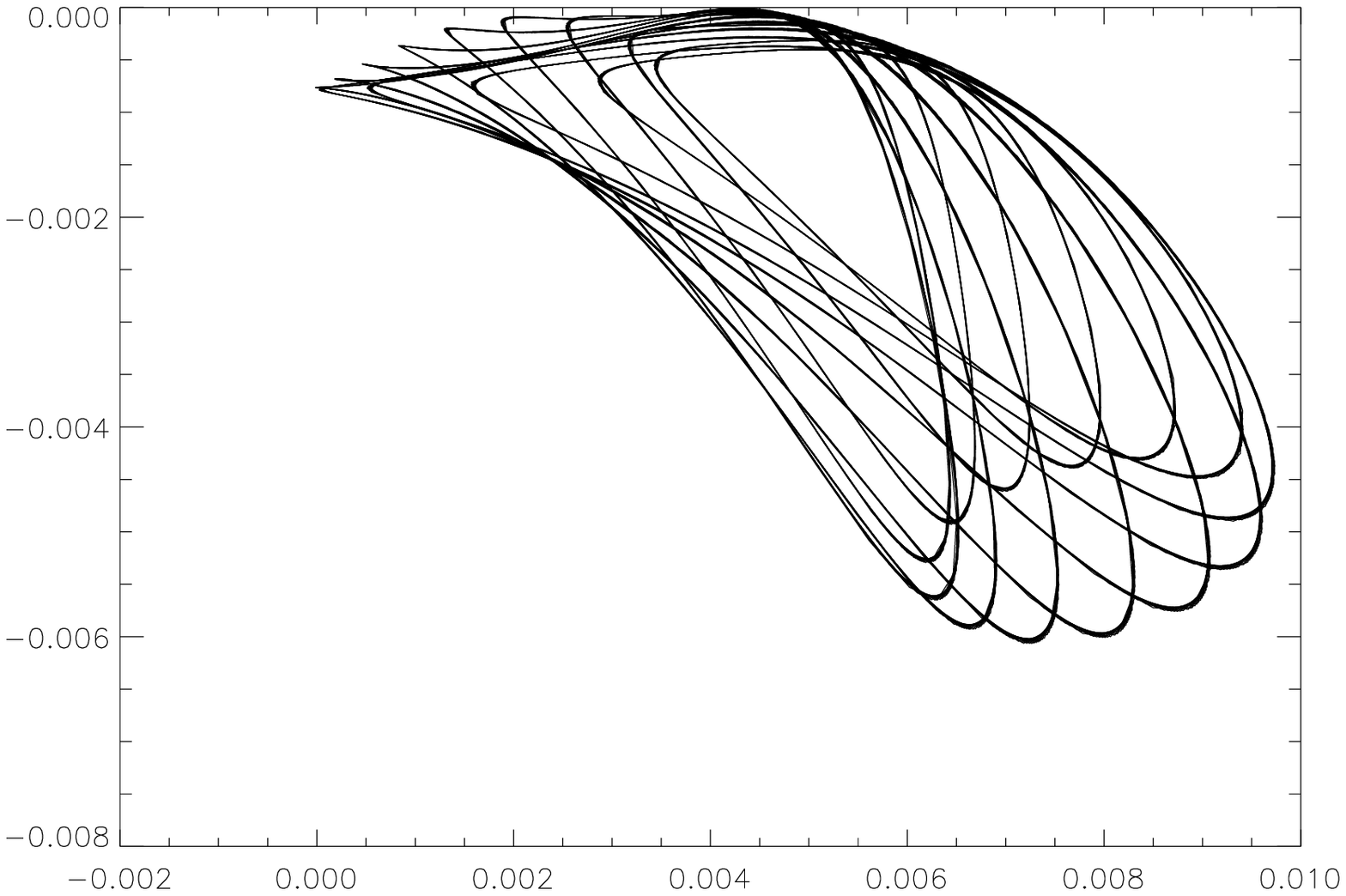,width=13cm,clip=}}
\vspace*{-5mm}
\mi
Figure 12b.
Projection of the trajectory in saturated regime (torus $T_{3,i}$)
on the plane of Fourier coefficients Im$\,b^1_{0,1,2}$
(horizontal axis) and Re$\,v^1_{0,1,2}$ (vertical axis) for $R=4$ and
$R_m=52$ (same run as on Fig.~11a).
\end{figure}

\pagebreak
\begin{figure}

\centerline{\psfig{file=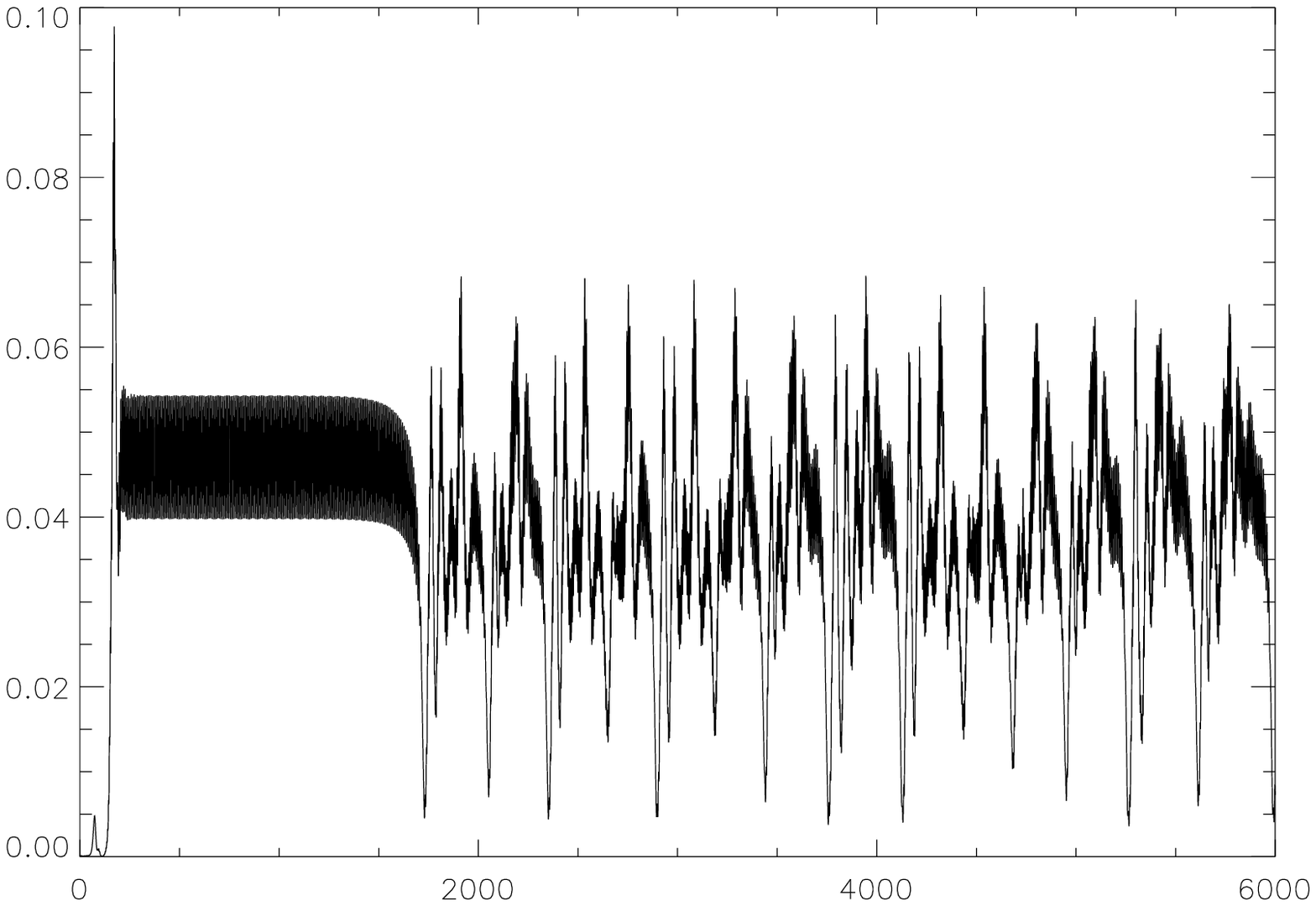,width=13cm,clip=}}
\vspace*{-5mm}
\mi
Figure 13.
Magnetic energy (vertical axis) as a function of time (horizontal axis)
for $R=15$ and $R_m=40$.

\vspace*{10mm}
\centerline{\psfig{file=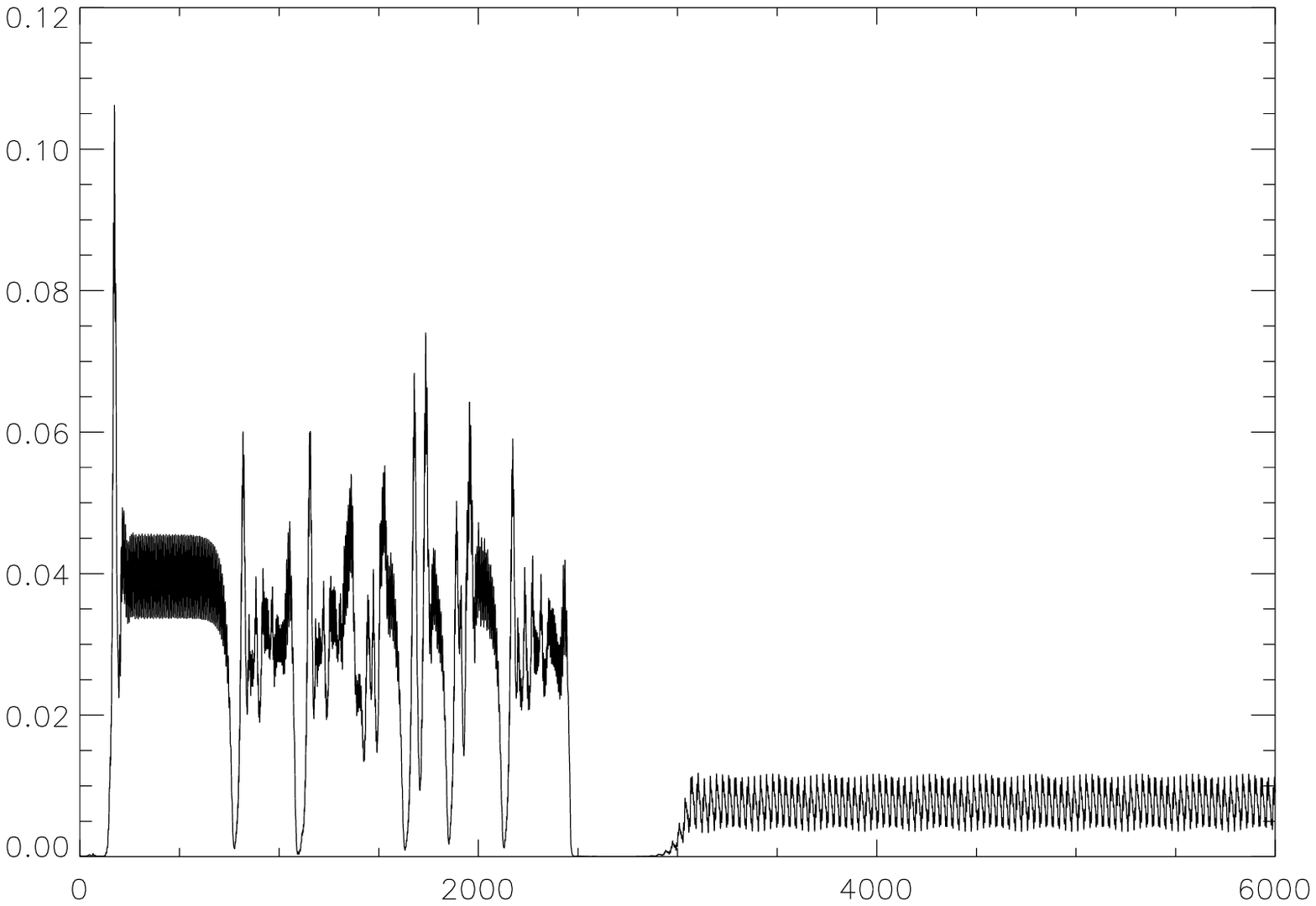,width=13cm,clip=}}
\vspace*{-5mm}
\mi
Figure 14.
Magnetic energy (vertical axis) as a function of time (horizontal axis)
for $R=20$ and $R_m=40$.
\end{figure}

\pagebreak
\begin{figure}

\centerline{\psfig{file=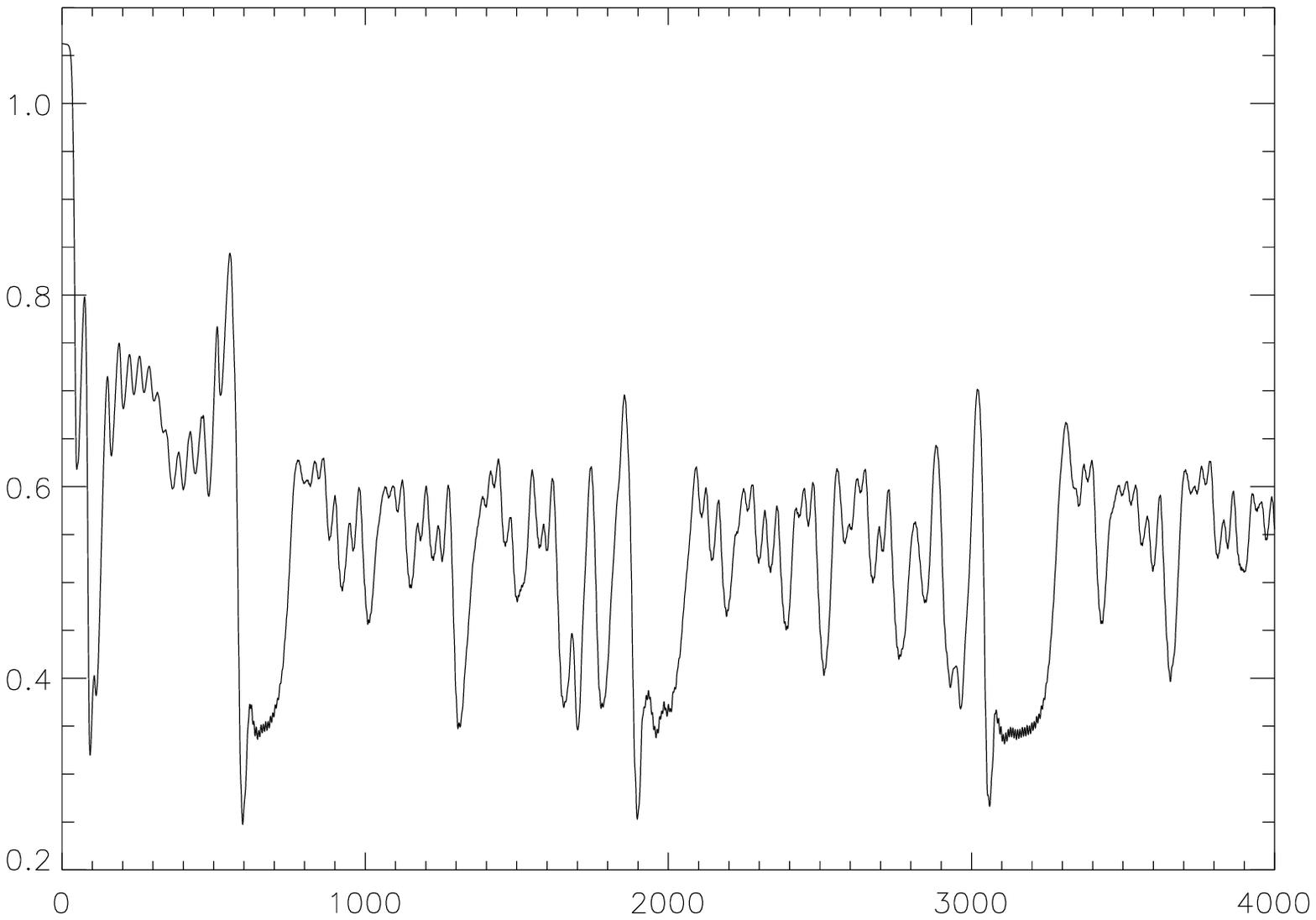,width=13cm,clip=}}
\vspace*{-5mm}
\mi
Figure 15a.
Kinetic energy (vertical axis) as a function of time (horizontal axis)
for $R=25$ and $R_m=40$.

\vspace*{10mm}
\centerline{\psfig{file=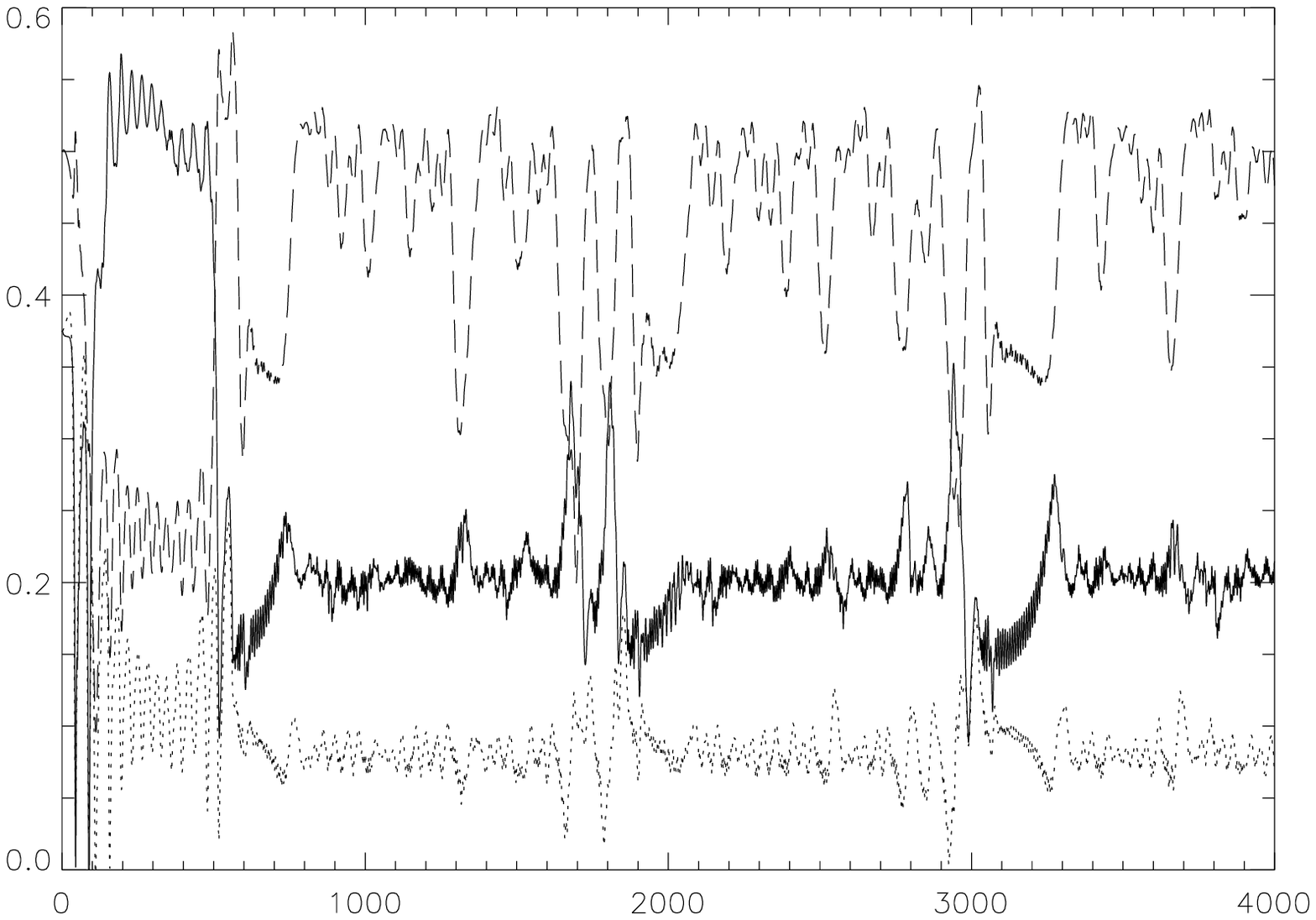,width=13cm,clip=}}
\vspace*{-5mm}
\mi
Figure 15b.
Fourier coefficients of the flow (vertical axis:
solid line -- Re$\,v_{1,0,0}^3$, dashed line -- Re$\,v_{0,1,0}^1$,
dot line -- Re$\,v_{0,0,1}^2$) as a function of time
(horizontal axis) for $R=25$ and $R_m=40$ (same run as on Fig.~14a).
\end{figure}

\pagebreak
\begin{figure}

\centerline{\psfig{file=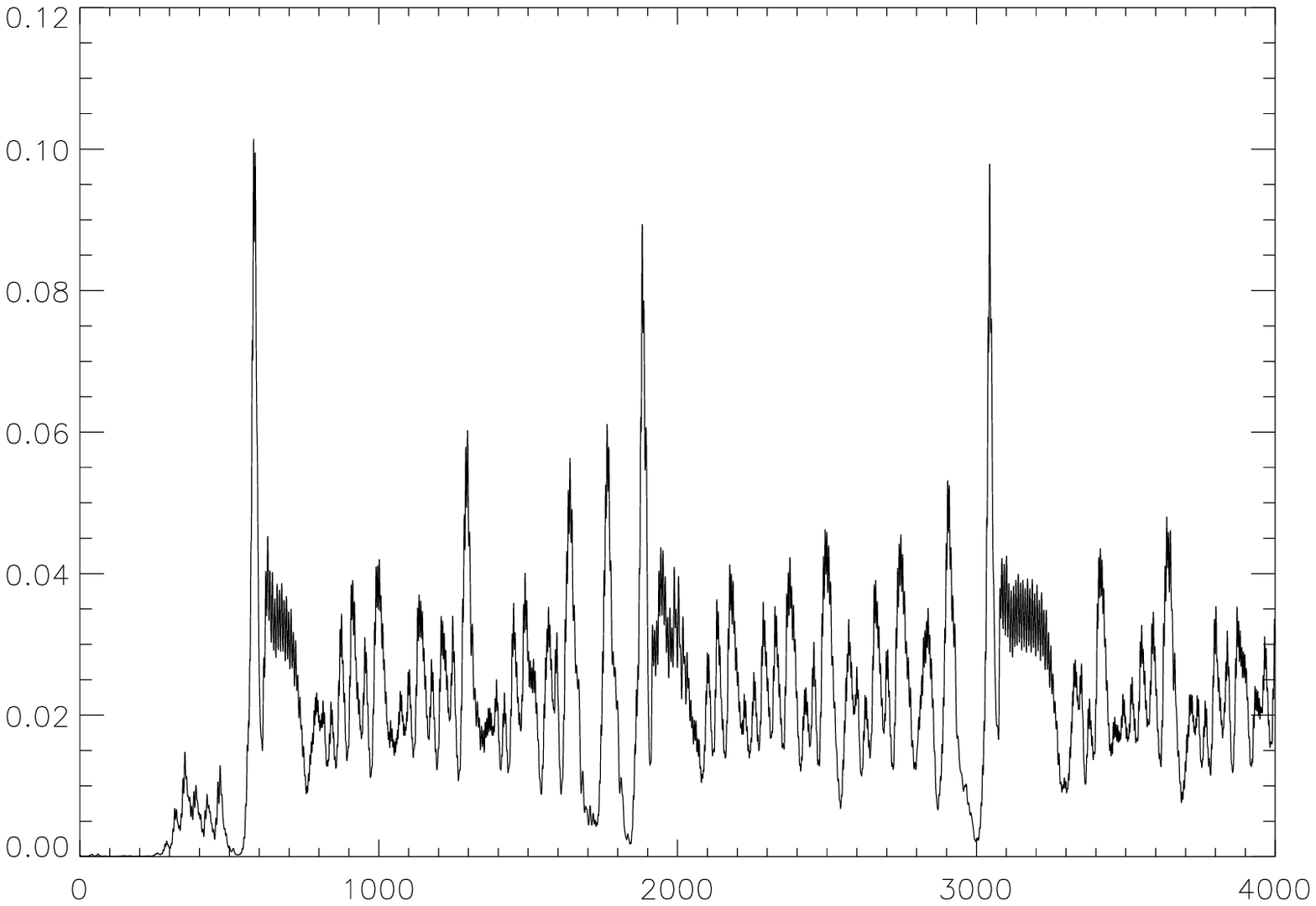,width=13cm,clip=}}
\vspace*{-5mm}
\mi
Figure 15c.
Magnetic energy (vertical axis) as a function of time (horizontal axis)
for $R=25$ and $R_m=40$ (same run as on Fig.~14a).

\vspace*{10mm}
\centerline{\psfig{file=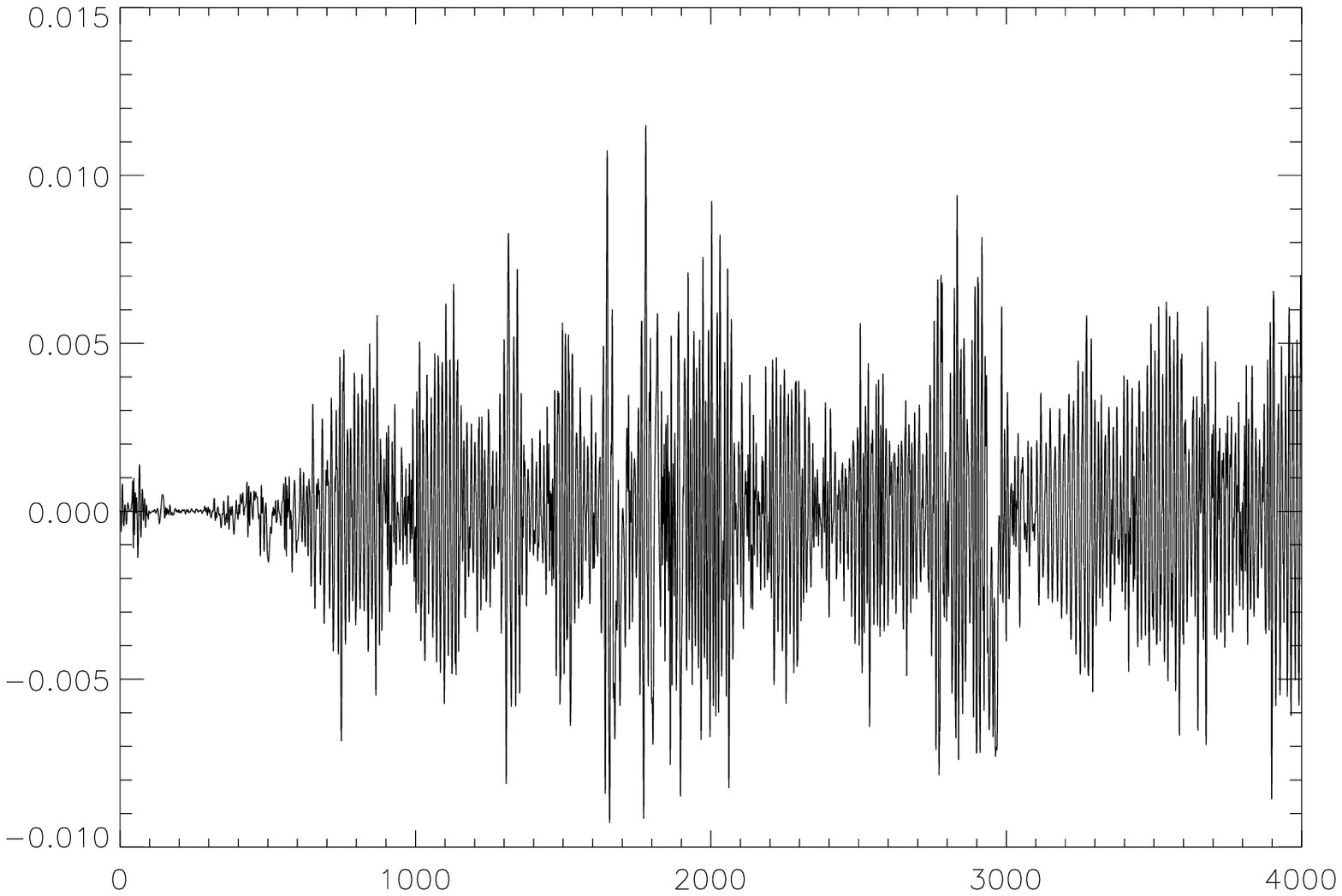,width=13cm,clip=}}
\vspace*{-5mm}
\mi
Figure 15d.
Fourier coefficient Im$\,b^1_{0,1,2}$ (vertical axis) as a function of time
(horizontal axis) for $R=25$ and $R_m=40$ (same run as on Fig.~14a).
\end{figure}

\pagebreak
\begin{figure}

\centerline{\psfig{file=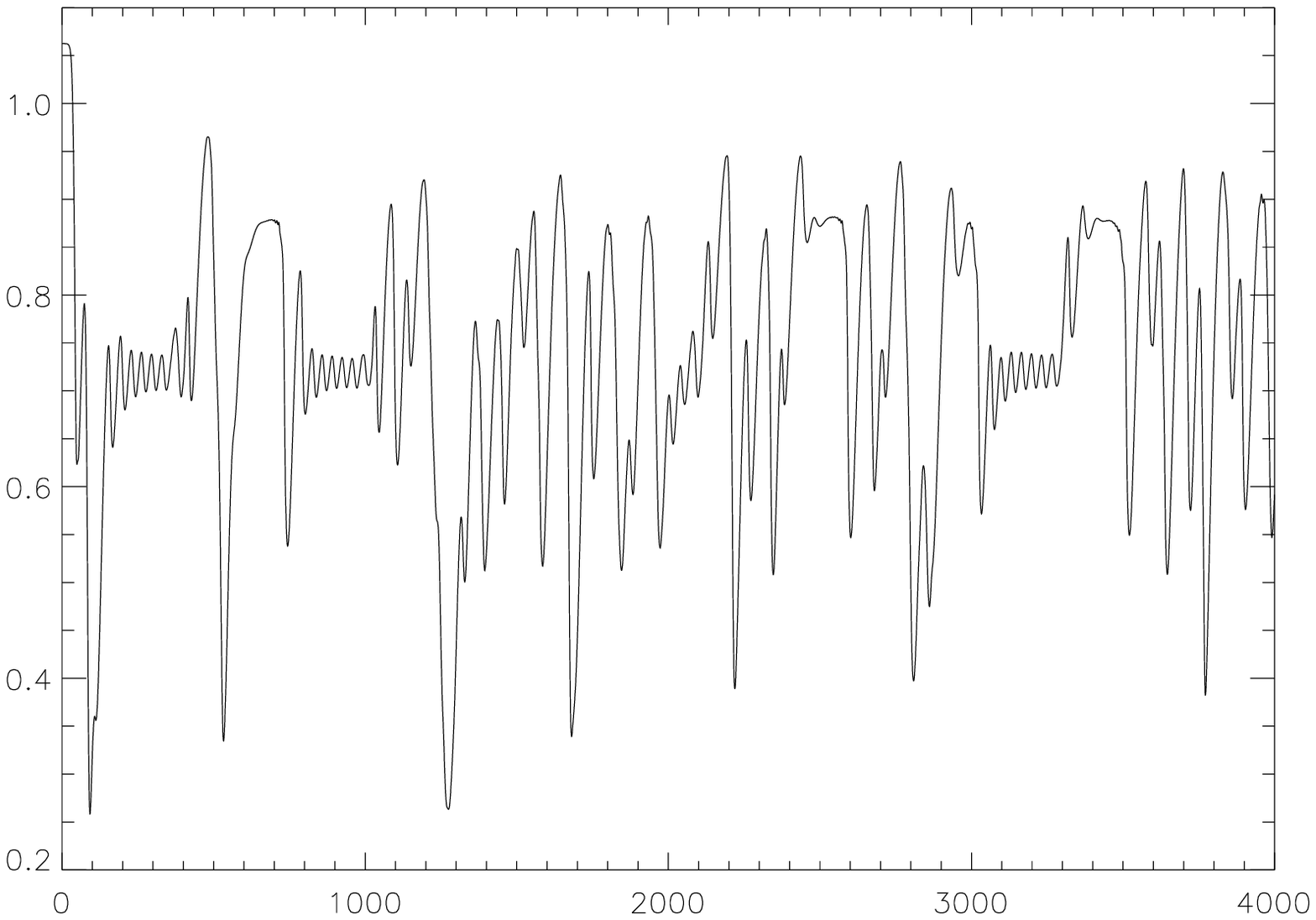,width=13cm,clip=}}
\vspace*{-5mm}
\mi
Figure 16a.
Kinetic energy (vertical axis) as a function of time (horizontal axis)
for $R=25$ in the hydrodynamic problem (the initial condition for the flow
is the same as on Fig.~14).

\vspace*{10mm}
\centerline{\psfig{file=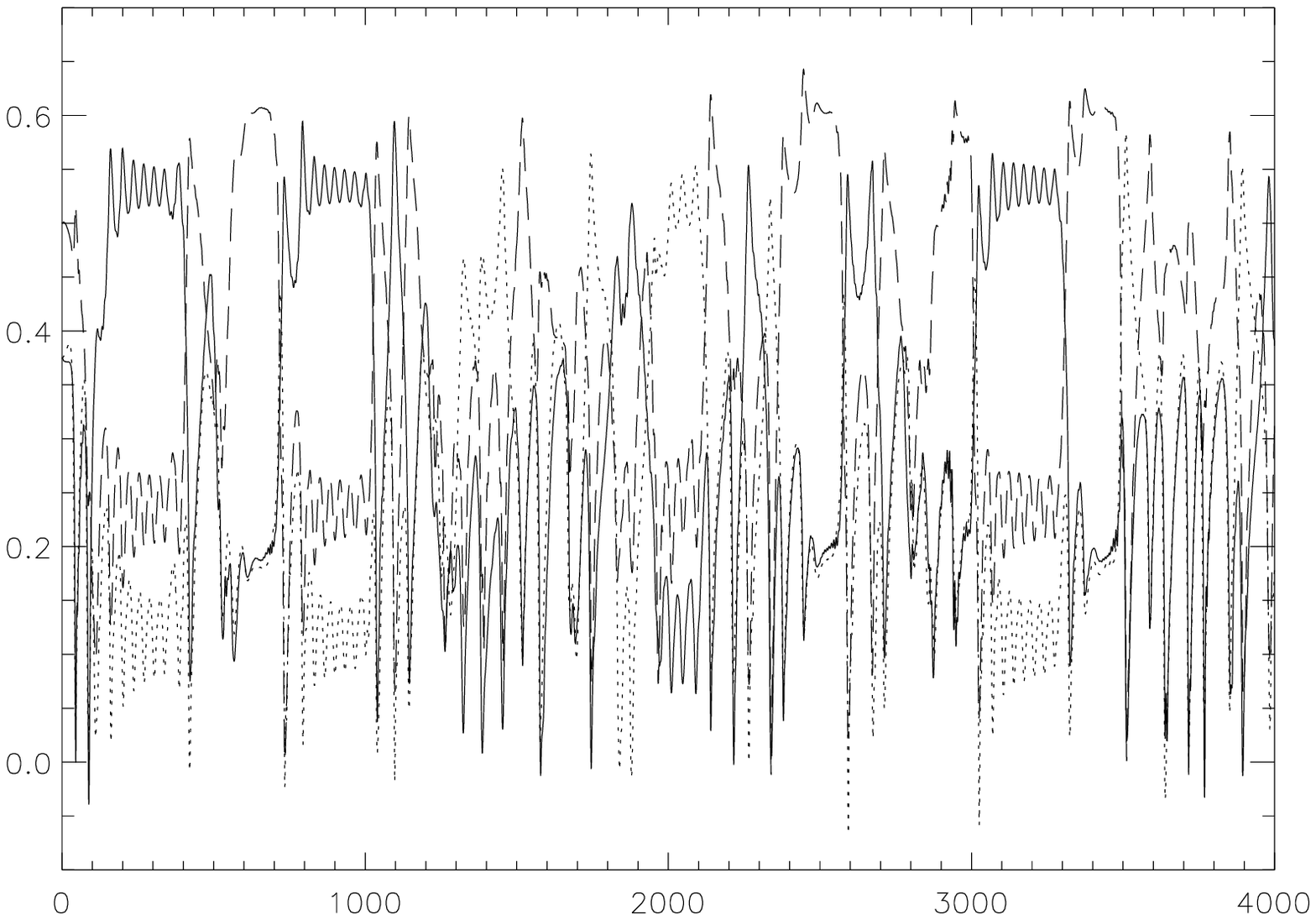,width=13cm,clip=}}
\vspace*{-5mm}
\mi
Figure 16b.
Fourier coefficients of the flow (vertical axis:
solid line -- Re$\,v_{1,0,0}^3$, dashed line -- Re$\,v_{0,1,0}^1$,
dot line -- Re$\,v_{0,0,1}^2$) as a function of time
(horizontal axis) for $R=25$ (same run as on Fig.~15a).
\end{figure}

\end{document}